\documentclass[twocolumn]{svjour3}
\usepackage{cite}

\usepackage[utf8]{inputenc}
\usepackage[english]{babel}
\usepackage[T1]{fontenc}

\usepackage{amsmath}
\usepackage{balance} 
\usepackage{amssymb}
\usepackage{pifont}
\newcommand{\xmark}{\ding{55}}%

\usepackage{stix}

\usepackage{booktabs} 

\usepackage[ruled]{algorithm2e} 

\SetAlFnt{\small}
\SetAlCapNameFnt{\small}
\SetAlCapHSkip{0pt}
\IncMargin{-\parindent}

\smartqed

\usepackage{float}

\usepackage{hyperref}

\usepackage{subcaption}
\usepackage[table]{xcolor}
\usepackage{comment}

\definecolor{Gray}{gray}{0.85}
\definecolor{LightGray}{gray}{0.75}
\newcolumntype{a}{>{\columncolor{Gray}}c}
\newcolumntype{b}{>{\columncolor{LightGray}}c}
\newcolumntype{d}{>{\columncolor{LightGray}}l}
\newcolumntype{e}{>{\columncolor{Gray}}l}

\newcommand{\para}[1]{\vspace{2mm}\noindent\textbf{#1}}

\newcounter{VasiaNOC}
\stepcounter{VasiaNOC}
\newcommand{\vasia}[1]{\textcolor{magenta}{\small \bf Vasia\#\arabic{VasiaNOC}\stepcounter{VasiaNOC}: #1}}

\usepackage{graphicx}
\newcommand{\stitle}[1]{\vspace{1ex}\noindent\textbf{#1}}

\newcounter{MariosNOC}
\stepcounter{MariosNOC}

\newcounter{AsteriosNOC}
\stepcounter{AsteriosNOC}

\newcounter{ParisNOC}
\stepcounter{ParisNOC}

\newcounter{mustCounter}
\stepcounter{mustCounter}

\newcounter{couldCounter}
\stepcounter{couldCounter}

\newcounter{shouldCounter}
\stepcounter{shouldCounter}

\begin{document}
\title{A Survey on the Evolution of Stream Processing Systems}

\author{Marios Fragkoulis$^{s}$  \and
        Paris Carbone$^{\dagger}$  \and Vasiliki Kalavri$^{\ddagger}$ \and Asterios Katsifodimos$^{\ast}$}

\date{}

\institute{$^{s}$Delivery Hero SE (marios.fragkoulis@deliveryhero.com), \\ $^{\ast}$Delft University of Technology (a.katsifodimos@tudelft.nl), \\ $^{\dagger}$RISE, parisc@kth.se - KTH (paris.carbone@ri.se)\\
$^{\ddagger}$Boston University (vkalavri@bu.edu)}

\maketitle

\sloppy 

\begin{abstract}
Stream processing has been an active research field for more than 20 years, but it is now witnessing its prime time due to recent successful efforts by the research community and numerous worldwide open-source communities. This survey provides a comprehensive overview of fundamental aspects of stream processing systems and their evolution in the functional areas of out-of-order data management, state management, fault tolerance, high availability, load management, elasticity, and reconfiguration. We review noteworthy past research findings, outline the similarities and differences between the  $1^{st}$ ('00-'10) and $2^{nd}$ ('11-'22) generation of stream processing systems, and discuss future trends and open problems.
\end{abstract}




\section{Introduction}

Applications of stream processing technology have gone through a resurgence, penetrating multiple and very diverse industries. Nowadays, virtually all Cloud vendors offer first-class support for deploying managed stream processing pipelines, while streaming systems are used in a variety of use-cases that go beyond the classic streaming analytics (windows, aggregates, joins, etc.). For instance, web companies are using stream processing for dynamic car-trip pricing, banks apply it for credit card fraud detection, while traditional industries apply streaming technology for real-time harvesting analytics. 
At the moment of writing we are witnessing a trend towards using stream processors to build more general event-driven architectures \cite{kleppmann2019online}, large-scale continuous ETL and analytics, and microservices \cite{operational-stream-processing}.

\begin{figure*}[t]
\centering
\includegraphics[width=.75\linewidth]{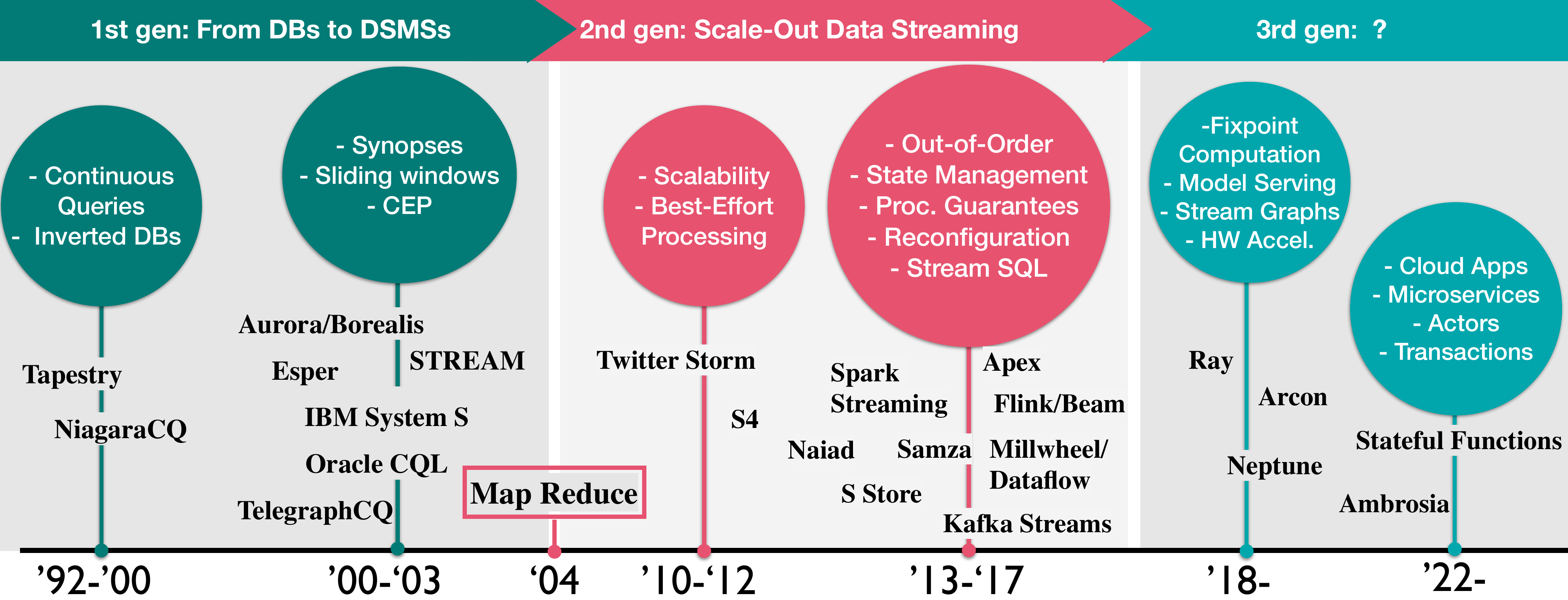}
\caption{\small An overview of the evolution of stream processing and respective domains of focus.}
\label{fig:overview}
\end{figure*}

During the last 20 years, streaming technology has evolved significantly, under the influence of database and distributed systems. The notion of streaming queries was first introduced in 1992 by the Tapestry system~\cite{terry1992continuous}, and was followed by lots of research on stream processing in the early 00s. Fundamental concepts and ideas originated in the database community and were implemented in prototypical systems such as TelegraphCQ~\cite{chandrasekaran2003telegraphcq}, Stanford's STREAM, NiagaraCQ~\cite{chen2000niagaracq}, Auroral/Borealis~\cite{abadi2003aurora}, and Gigascope~\cite{cranor2003gigascope}. Although these prototypes roughly agreed on the data model, they differed considerably on querying semantics \cite{cql,secret}. This research period also introduced various systems challenges, such as sliding window aggregation \cite{arasu2004resource,li2005no}, fault-tolerance and high-availability \cite{shah2003flux,balazinska2008fault}, as well as load balancing and shedding \cite{tatbul2003load}. This first wave of research was highly influential to commercial stream processing systems that were developed in the following years (roughly during 2004 -- 2010), such as IBM System S, Esper, Oracle CQL/CEP and TIBCO. These systems focused -- for the most part -- on streaming window queries and Complex Event Processing (CEP). This era of systems was mainly characterized by scale-up architectures, processing ordered event streams.

The second generation of streaming systems was a result of research that started roughly after the introduction of MapReduce~\cite{dean2004mapreduce} and the popularization of Cloud Computing. The focus shifted towards distributed, data-parallel processing engines and shared-nothing architectures on commodity hardware. Lacking well-defined semantics and a proper query language, systems like Millwheel \cite{akidau2013millwheel}, Storm~\cite{CUSTOM:web/Storm}, Spark Streaming~\cite{zaharia2012discretized}, and Apache Flink~\cite{carbone2015apache} first exposed primitives for expressing streaming computations as hard-coded dataflow graphs and transparently handled data-parallel execution on distributed clusters. With very high influence, the Google Dataflow model \cite{akidau2015dataflow} re-introduced older ideas such as out-of-order processing \cite{li2008out} and punctuations~\cite{Tucker2003Exploiting}, proposing a unified parallel processing model for streaming and batch computations. Stream processors of this era are converging towards fault-tolerant, scale-out processing of massive out-of-order streams.

Figure~\ref{fig:overview} presents a schematic categorization of influential streaming systems into three generations and highlights each era's domains of focus. Although the foundations of stream processing have remained largely unchanged over the years, stream processing systems have transformed into sophisticated and scalable engines, producing correct results in the presence of failures. Early systems and languages were designed as extensions of relational execution engines, with the addition of windows. Modern streaming systems have evolved in the way they reason about completeness and ordering (e.g., out-of-order computation) and have witnessed architectural paradigm shifts that constituted the foundations of processing guarantees, reconfiguration, and state management. At the moment of writing, we observe yet another paradigm shift towards general event-driven architectures, actor-like programming models and microservices~\cite{bykov2011orleans,akhter2019stateful}, and a growing use of modern hardware~\cite{koliousis2016saber,zhang2020hardware,thomas2020fleet,zeuch2019analyzing}.

This survey is the first to focus on the evolution of streaming systems rather than the state of the field at a particular point in time. To the best of our knowledge, this is also the first attempt at understanding the underlying reasons why certain early techniques and designs prevailed in modern systems while others were abandoned. Further, by examining how ideas survived, evolved, and were often re-invented, we reconcile the terminology used by the different generations of streaming systems.



\subsection{Contributions}
With this survey paper, we make the following contributions:
\begin{itemize}
    \item We summarize existing approaches to streaming systems design and categorize early and modern stream processors in terms of underlying assumptions and mechanisms.
    \item We compare early and modern stream processing systems with regard to out-of-order data management, state management, fault-tolerance, high availability, load management, elasticity, and reconfiguration.
    \item We highlight important but overlooked works that have influenced today's streaming systems design.
    \item We establish a common nomenclature for fundamental streaming concepts, often described by inconsistent terms in different systems and communities.
\end{itemize}

\subsection{Related surveys and research collections}

We view the following surveys as complementary to ours and recommend them to readers interested in diving deeper into a particular aspect of stream processing or those who seek a comparison between streaming technology and advances from adjacent research communities.

\setlength{\tabcolsep}{8pt}
\begin{table*}[ht]
\centering
\caption{Evolution of streaming systems}
\label{tab:evolution-streaming}
\begin{tabular}{p{0.3\textwidth}
                        p{0.4\textwidth}
                        p{0.4\textwidth}
                        }
\hline
\multicolumn{1}{l}{\textbf{}} &
\multicolumn{1}{l}{\textbf{1st generation}} &
\multicolumn{1}{l}{\textbf{2nd-3rd generation}}
\\
\hline


\multicolumn{1}{l}{\textbf{Results}}
& \multicolumn{1}{e}{approximate or exact}
& \multicolumn{1}{d}{exact}
\\

\multicolumn{1}{l}{\textbf{Language}}
& \multicolumn{1}{e}{SQL extensions, CQL}
& \multicolumn{1}{d}{UDF-heavy -- Java, Scala, Python, SQL-like, etc.}
\\

\multicolumn{1}{l}{\textbf{Query plans}}
& \multicolumn{1}{e}{global, optimized, with pre-defined operators}
& \multicolumn{1}{d}{independent, with custom operators}
\\

\multicolumn{1}{l}{\textbf{Execution}}
& \multicolumn{1}{e}{(mostly) scale-up}
& \multicolumn{1}{d}{distributed}
\\

\multicolumn{1}{l}{\textbf{Parallelism}}
& \multicolumn{1}{e}{pipeline}
& \multicolumn{1}{d}{data, pipeline, task}
\\

\multicolumn{1}{l}{\textbf{Time \& progress}}
& \multicolumn{1}{e}{heartbeats, slack, punctuations}
& \multicolumn{1}{d}{low-watermark, frontiers}
\\

\multicolumn{1}{l}{\textbf{State management}}
& \multicolumn{1}{e}{shared synopses, in-memory}
& \multicolumn{1}{d}{per query, partitioned, persistent, larger-than-memory}
\\

\multicolumn{1}{l}{\textbf{Fault tolerance}}
& \multicolumn{1}{e}{HA-focused, limited  correctness guarantess}
& \multicolumn{1}{d}{distributed snapshots, exactly-once}
\\

\multicolumn{1}{l}{\textbf{Load management}}
& \multicolumn{1}{e}{load shedding, load-aware scheduling}
& \multicolumn{1}{d}{backpressure, elasticity}
\\
\hline

\end{tabular}
\end{table*}

Cugola and Margara~\cite{Cugola2012Flows} provide a view of stream processing with regard to related technologies, such as active databases and complex event processing systems, and discuss their relationship with data streaming systems. Further, they provide a categorization of streaming languages and streaming operator semantics. The language aspect is further covered in another recent survey~\cite{Hirzel2018Languages}, which focuses on the languages developed to address the challenges in very large data streams. It characterizes  streaming languages in terms of data model, execution model, domain, and intended user audience. R{\"o}ger and Mayer~\cite{Henriette2019Elasticity} present an overview of recent work on parallelization and elasticity approaches of streaming systems. They define a general system model which they use to introduce operator parallelization strategies and parallelism adaptation methods. Their analysis also aims at comparing elasticity approaches originating in different research communities. Hirzel et al.~\cite{Hirzel2014Catalog} present an extensive list of logical and physical optimizations for streaming query plans. They present a categorization of streaming optimizations in terms of their assumptions, semantics, applicability scenarios, and trade-offs. They also present experimental evidence to reason about profitability and guide system implementers in selecting appropriate optimizations. To, Soto, and Markl~\cite{To2017State} survey the concept of state and its applications in big data management systems, covering also aspects of streaming state.
Finally, Dayarathna and Perera~\cite{Dayarathna2018Recent} present a survey of the advances of the last decade with a focus on system architectures, use-cases, and hot research topics. They summarize recent systems in terms of their features, such as what types of operations they support, their fault-tolerance capabilities, their use of programming languages, and their best reported performance.

Theoretical foundations of streaming data management and streaming algorithms are out of the scope of this survey. A comprehensive collection of influential works on these topics can be found in Garofalakis et al.~\cite{garofalakis2016data}. The collection focuses on major contributions of the first generation of streaming systems. It reviews basic algorithms and synopses, fundamental results in stream data mining, streaming languages and operator semantics, and a set of representative applications from different domains.

\subsection{Survey organization}
We begin by presenting the essential elements of the domain in Section~\ref{sec:preliminaries}.
Then we elaborate on each of the important functionalities offered by stream processing
systems: out-of-order data management (Section~\ref{sec:order}), state management (Section~\ref{sec:state}),
fault tolerance and high availability (Section~\ref{sec:FT}), and load management, elasticity, and reconfiguration (Section~\ref{sec:elasticity}). Each one of these sections contains a \emph{Vintage vs. Modern} discussion that compares early to contemporary approaches and a summary of open problems.
We summarize our major findings, discuss prospects, and conclude in Table~\ref{tab:evolution-streaming} and Section~\ref{sec:conclusion}.

\section{Preliminaries}
\label{sec:preliminaries}

In this section, we provide necessary background and explain fundamental stream processing concepts the rest of this survey relies on. We discuss the key requirements of a streaming system, introduce the basic streaming data models, and give a high-level overview of the architecture of early and modern streaming systems.

\subsection{Requirements of streaming systems}\label{sec:requirements}
A data stream is a data set that is produced incrementally over time, rather than being available in full before its processing begins~\cite{garofalakis2016data}. Data streams are high-volume, real-time data that might be unbounded. Therefore, stream processing systems can neither store the entire stream in an accessible way nor can they control the data arrival rate or order. In contrast to traditional data management infrastructure, streaming systems have to process elements on-the-fly using limited memory. Stream elements arrive continuously and either bear a timestamp or are assigned one on arrival.

Respectively, a streaming query ingests events and produces results in a continuous manner, using a single pass or a limited number of passes over the data. Streaming query processing is challenging for multiple reasons. First, continuously producing updated results might require storing historical information about the stream seen so far in a compact representation that can be queried and updated efficiently. Such summary representations are known as \emph{sketches} or \emph{synopses}. Second, in order to handle high input rates, certain queries might not afford to continuously update indexes and materialized views. Third, stream processors cannot rely on the assumption that state can be reconstructed from associated inputs. To achieve acceptable performance, streaming operators need to leverage incremental computation.

The aforementioned characteristics of data streams and continuous queries provide a set of unique requirements for streaming systems, other than the evident performance ones of low latency and high throughput. Given the lack of control over the input order, a streaming system needs to produce correct results when receiving out-of-order and delayed data (cf. Section~\ref{sec:order}). It needs to implement mechanisms that estimate a stream's progress and reason about result completeness. Further, the long-running nature of streaming queries demands that streaming systems manage accumulated state (cf. Section~\ref{sec:state}) and guard it against failures (cf. Section~\ref{sec:FT}). Finally, having no control over the data input rate requires stream processors to be adaptive so that they can handle workload variations without sacrificing performance (cf. Section~\ref{sec:elasticity}).


\subsection{Streaming data models}
There exist many theoretical streaming data models, mainly serving the purpose of studying the space requirements and computational complexity of streaming algorithms and understanding which streaming computations are practical.
For instance, a stream can be modeled as a dynamic one-dimensional vector~\cite{garofalakis2016data}. The model defines how this dynamic vector is updated when a new element of the stream becomes available. While theoretical streaming data models are useful for algorithm design, early stream processing systems instead adopted extensions of the \emph{relational} data model. Recent streaming dataflow systems, especially those influenced by the MapReduce philosophy, place the responsibility of data stream modeling on the application developer.

\subsubsection{Relational Streaming Model}
In the relational streaming model as implemented by first-generation systems~\cite{arasu2003stream, chandrasekaran2003telegraphcq, abadi2003aurora, cranor2003gigascope}, a stream is interpreted as describing a changing relation over a common schema. Streams are either produced by external sources and update relation tables or are produced by continuous queries and update materialized views. An operator outputs event streams that describe the changing view computed over the input stream according to the relational semantics of the operator. Thus, the semantics and schema of the relation are imposed by the system. 

STREAM~\cite{arasu2003stream} defines streams as bags of tuple-timestamp pairs and relations as time-varying bags of tuples. The implementation unifies both types as sequences of timestamped tuples, where each tuple also carries a flag that denotes whether it is an insertion or a deletion. Input streams consist of insertions only, while relations may also contain deletions. TelegraphCQ~\cite{chandrasekaran2003telegraphcq} uses a similar data model. Aurora~\cite{abadi2003aurora} models streams as append-only sequences of tuples, where a set of attributes denote the key and the rest of the attributes denote values. Borealis~\cite{abadi2005design} generalizes this model to support insertion, deletion, and replacement messages. Messages may also contain additional fields related to QoS metrics. Gigascope~\cite{cranor2003gigascope} extends the sequence database model. It assumes that stream elements bear one or more timestamps or sequence numbers, which generally increase (or decrease) with the ordinal position of a tuple in a stream. Ordering attributes can be (strictly) monotonically increasing or decreasing, monotone non-repeating, or increasing within a group of records. In CEDR~\cite{Barga2007Consistent}, stream elements bear a valid timestamp, $V_{s}$, after which they are considered valid and can contribute to the result. Alternatively, events can have validity intervals. The contents of the relation at time $t$ are all events with $V_{s} \leq t$.

\subsubsection{Dataflow Streaming Model}
The dataflow streaming model, as implemented by systems of the second generation~\cite{carbone2015apache,zaharia2012discretized,akidau2015dataflow}, does not impose any strict schema or semantics to the input stream elements, other than the presence of a timestamp.
While some systems, like Naiad~\cite{murray2013naiad}, require that all stream elements bear a logical timestamp, other systems, such as Flink~\cite{carbone2015apache} and Dataflow~\cite{akidau2015dataflow}, expect the declaration of a \emph{time domain}. Applications can operate in one of three modes: (i) \emph{event} (or application) time is the time when events are generated at the sources, (ii) \emph{processing} time is the time when events are processed in the streaming system, and (iii) \emph{ingestion} time is the time when events arrive at the system.
Modern dataflow streaming systems can ingest any type of input stream, irrespectively of whether its elements represent additions, deletions, replacements or deltas. The application developer is responsible for imposing the semantics and writing the operator logic to update state accordingly and produce correct results. Designating keys and values is also usually not required at ingestion time, however, keys must be defined when using certain data-parallel operators, such as windows.

\subsection{Architectures of Streaming Systems}
The general architecture of streaming systems has evolved significantly over the last two decades. 
Before we delve into the specific approaches to out-of-order management, state, fault tolerance, and load management, 
we outline some fundamental differences between early (1st generation) and modern (2nd generation) streaming systems. Table \ref{tab:evolution-streaming} summarizes our findings.

The architecture of a first-generation DSMS follows closely that of a database management systems (DBMS), with the addition of certain components designated to address the requirements of streaming data (cf. Section~\ref{sec:requirements}). In particular, the input manager is responsible for ingesting streams and possibly buffering and ordering input elements.
The scheduler determines the order or operator execution, as well as the number of tuples to process and push to the outputs. Two important additional components are the quality monitor and load shedder which monitor stream input rates and query performance and selectively drop input records to meet target latency requirements. Queries are compiled into a shared query plan which is optimized and submitted to the query execution engine. In the common case, a DSMS supports both ad-hoc and continuous queries. Early architectures are designed with the goal to provide fast, but possibly approximate results to queries.

The next generation of distributed dataflow systems are usually deployed on shared-nothing clusters of machines. Dataflow systems employ task and data parallelism, have explicit state management support, and implement advanced fault-tolerance capabilities to provide result guarantees. Distributed workers execute parallel instances of one of more operators (tasks) on disjoint stream partitions. In contrast to DSMSs, queries are independent of each other, maintain their own state, and they are assigned dedicated resources. Every query is configured individually and submitted for execution as a separate job. Input sources are typically assumed to be replayble and state is persisted to embedded or external stores. Modern architectures prioritize high throughput, robustness, and result correctness over low latency.

Despite the evident differences between early and modern streaming systems' architectures, many fundamental aspects have remained unchanged in the past two decades. The following sections examine in detail how streaming systems have evolved in terms of out-of-order processing, state capabilities, fault-tolerance, and load management.

\section{Managing Event Order and Timeliness}\label{sec:order}
A streaming system receives data continuously from one or more input sources.
Typically the order of data in a stream is part of the stream's semantics~\cite{maier2005semantics}.
Depending on the computations to perform, a streaming system may have to process stream tuples in a certain order to provide semantically correct results~\cite{srivastava2004flexible}.
However, in the general case, a stream's data tuples arrive out of
order~\cite{lamport1978time,Tucker2003Exploiting} for reasons explained in Section~\ref{subsec:disorder-causes}.

\begin{description}
    \item \textit{Out-of-order} data tuples~\cite{srivastava2004flexible,Traub2019efficient} arrive in a streaming system after tuples with later event time timestamps.
\end{description}

In the rest of the paper we use the terms disorder~\cite{maier2005semantics} and out-of-order~\cite{li2008out, akidau2013millwheel} to refer to the disturbance of order in a stream's data tuples.
Reasoning about order and managing disorder are fundamental considerations for the operation of streaming systems.

In the following, we highlight the causes of disorder in Section~\ref{subsec:disorder-causes}, clarify the relationship between disorder in a stream's tuples and processing progress in Section~\ref{subsec:disorder-progress}, and outline the two key system architectures for managing out-of-order data in Section~\ref{subsec:disorder-architectures}.
Then, we describe the consequences of disorder in Section~\ref{subsec:disorder-effects} and present the mechanisms for managing disorder in Section~\ref{subsec:disorder-mechanisms}.
Finally, in Section~\ref{sub:vintage}, we discuss the differences of out-of-order data management in early and modern systems and we present open problems in Section~\ref{sub:open}.

\subsection{Causes of Disorder}
\label{subsec:disorder-causes}
Disorder in data streams may be owed to stochastic factors that are external to a streaming system or to the operations taking place inside the system.

The most common external factor that introduces disorder to streams is the network~\cite{srivastava2004flexible,Krishnamurthy2010Continuous}.
Depending on the network's reliability, bandwidth, and load, the routing of some stream tuples can take longer to complete compared to the routing of others, leading to a different arrival order in a streaming system.
Even if the order of tuples in an individual stream is preserved, ingestion from multiple sources,
such as sensors, typically results in a disordered collection of tuples, unless the sources are carefully coordinated, which is rare.

External factors aside, specific operations on streams break tuple order.
First, join processing takes two streams and produces a shuffled combination of the two, since a parallel join operator repartitions the data according to the join attribute~\cite{Urhan00XJoin} and outputs join results by order of match~\cite{kang2003evaluating,
 hammad2003scheduling}.
Second, windowing based on an attribute different to the ordering attribute reorders the stream~\cite{cranor2003gigascope}.
Third, data prioritization~\cite{Raman1999Online, Urhan2001Dynamic} by using an attribute different to the ordering one also changes the stream's order.
Finally, the union operation on two unsynchronized streams yields a stream with all tuples of the two input streams interleaving each other in random order~\cite{abadi2003aurora}.

\subsection{Disorder and Processing Progress}
\label{subsec:disorder-progress}

In order to manage disorder, streaming systems need to detect processing progress.
We discuss how disorder management and progress tracking are intertwined in Sections~\ref{subsec:disorder-architectures} and~\ref{subsec:disorder-effects}.

\emph{Progress} regards how much the processing of a stream’s tuples has advanced over time.
Processing progress can be defined and quantified with the aid of an attribute \textit{A} of a stream’s tuples that orders the stream.
The processing of the stream progresses when the smallest value of \textit{A} among the unprocessed tuples increases over time~\cite{li2008out}.
\textit{A} then is a \emph{progressing attribute} and the oldest value of \textit{A} per se, is a measure of progress because it denotes
how far in processing tuples the system has reached since the beginning.
Beyond this definition, streaming systems often make their own interpretation of progress, which may involve more than one attributes.

\subsection{System Architectures for Managing Disorder}
\label{subsec:disorder-architectures}



Two main architectural archetypes have influenced the design of streaming systems with respect to managing disorder: (i) in-order processing systems~\cite{cql, srivastava2004flexible, abadi2003aurora, cranor2003gigascope}, and (ii) out-of-order processing systems~\cite{li2008out, akidau2013millwheel, murray2013naiad, carbone2015apache}.

In-order processing systems manage disorder by fixing a stream's order. As a result, they essentially track progress by monitoring how far the processing of a data stream has advanced.
In-order systems buffer and reorder tuples up to a \emph{lateness} bound.
Then, they forward the reordered tuples for processing and clear the corresponding buffers.

In out-of-order processing systems, operators or a global authority produce progress information using any of the metrics detailed in Section~\ref{subsubsec:tracking-progress}, and propagate it to the dataflow graph.
The information typically reflects the oldest unprocessed tuple in the system and establishes a lateness bound for admitting out-of-order tuples.
In contrast to in-order systems, tuples are processed without delay in their arrival order, as long as they do not exceed the lateness bound.

\subsection{Effects of Disorder}
\label{subsec:disorder-effects}

In unbounded data processing, disorder can impede progress~\cite{li2008out} or lead to wrong results if ignored~\cite{srivastava2004flexible}.

Disorder affects processing progress when the operators that comprise the topology of the computation require ordered input.
Various implementations of \textit{join} and \textit{aggregate} rely on ordered input to produce correct results~\cite{abadi2003aurora, srivastava2004flexible}.
When operators in in-order systems receive out-of-order tuples, they have to reorder them prior to including them in the window they belong.
Reordering, however, imposes processing overhead, memory space overhead, and latency.
Out-of-order systems, on the other hand, track progress and process data in whatever order they arrive, up to the lateness bound. To include late tuples in results, they additionally need to store the processing state up to the lateness bound.
As a sidenote, order-insensitive operators~\cite{abadi2003aurora, srivastava2004flexible, li2008out}, such as \textit{apply, project, select, dupelim}, and \textit{union}, are agnostic to  disorder in a stream and produce correct results even when presented with disordered input.

Ignoring out-of-order data might lead to incorrect results if the output is computed on partial input only. Thus, a streaming system needs to be capable of processing out-of-order data and incorporate their effect to the computation.
However, without knowledge of how late data can be, waiting indefinitely can block output and accumulate large computation state.
This concern manifests on all architectures and we discuss how it can be countered with disorder management mechanisms, next. 

\subsection{Mechanisms for Managing Disorder}
\label{subsec:disorder-mechanisms}


In this section, we elaborate on influential mechanisms for managing disorder in unbounded data, namely slack~\cite{abadi2003aurora}, heartbeats~\cite{srivastava2004flexible}, low-watermarks~\cite{li2008out}, pointstamps~\cite{murray2013naiad}, and triggers~\cite{akidau2015dataflow}.
Heartbeats, low-watermarks, and pointstamps track processing progress and quantify a lateness bound using a metric, such as time.
In contrast, slack merely quantifies the lateness bound.
If tuples arrive after the lateness bound expires, triggers can be used to update computation results in \emph{revision processing}~\cite{abadi2005design}.
We also discuss punctuations~\cite{Tucker2003Exploiting}, a generic mechanism for communicating information across the dataflow graph, that has been heavily used as a vehicle in managing disorder.

\subsubsection{Tracking processing progress}
\label{subsubsec:tracking-progress}

\begin{figure}[ht]
    \centering
    \begin{subfigure}{.97\linewidth}
        \centering
        \includegraphics[width=.8\linewidth]{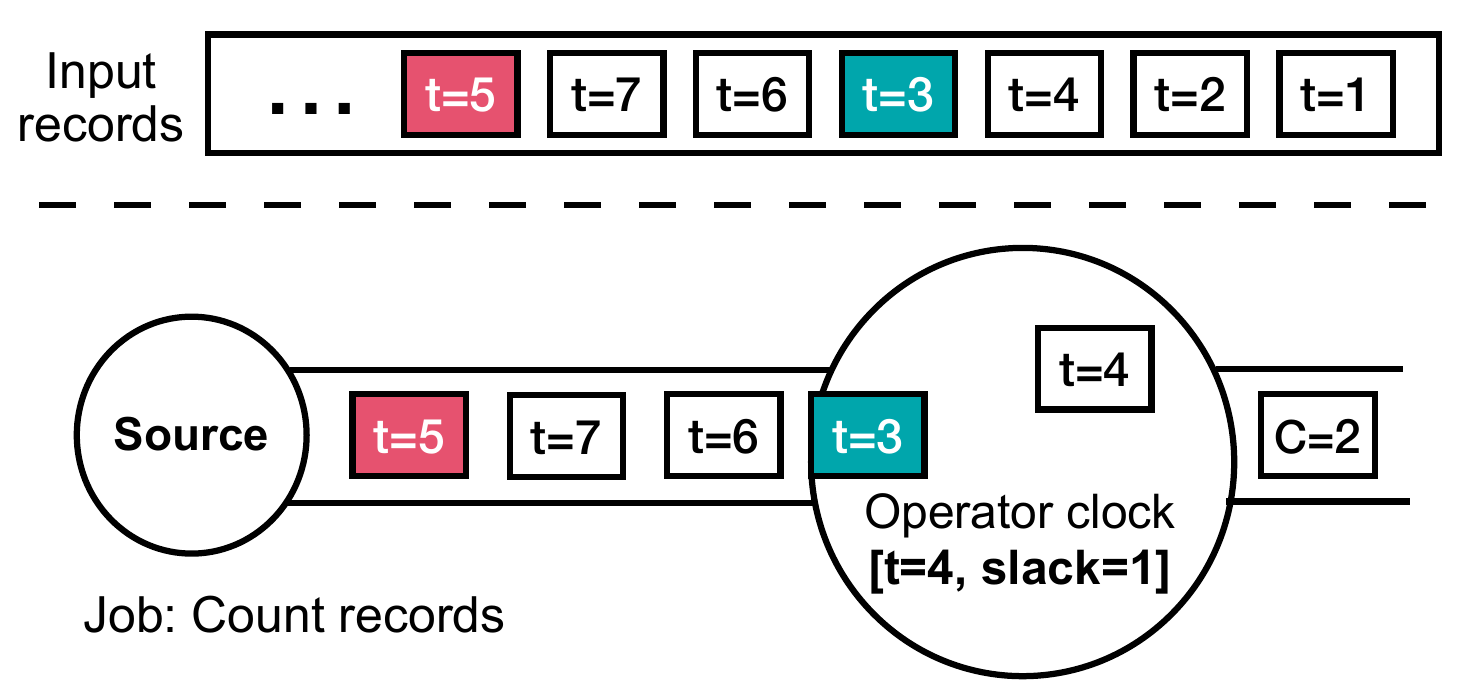}
        \caption{Slack}
        \label{fig:slack}
    \end{subfigure}
    \begin{subfigure}{.97\linewidth}
        \centering
        \includegraphics[width=.9\linewidth]{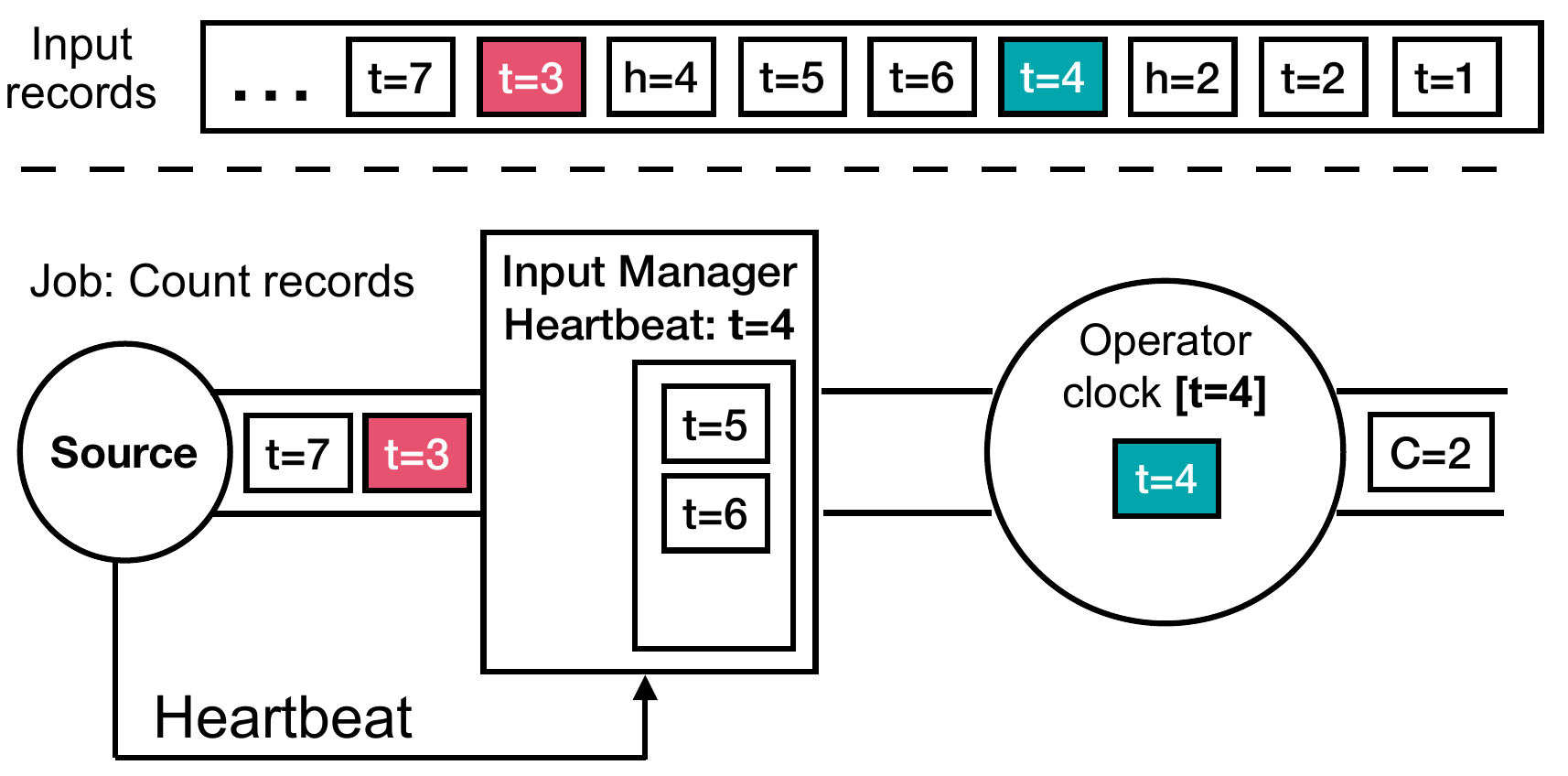}
        \caption{Heartbeat}
        \label{fig:heartbeat}
    \end{subfigure}
        \begin{subfigure}{.97\linewidth}
        \centering
        \includegraphics[width=.89\linewidth]{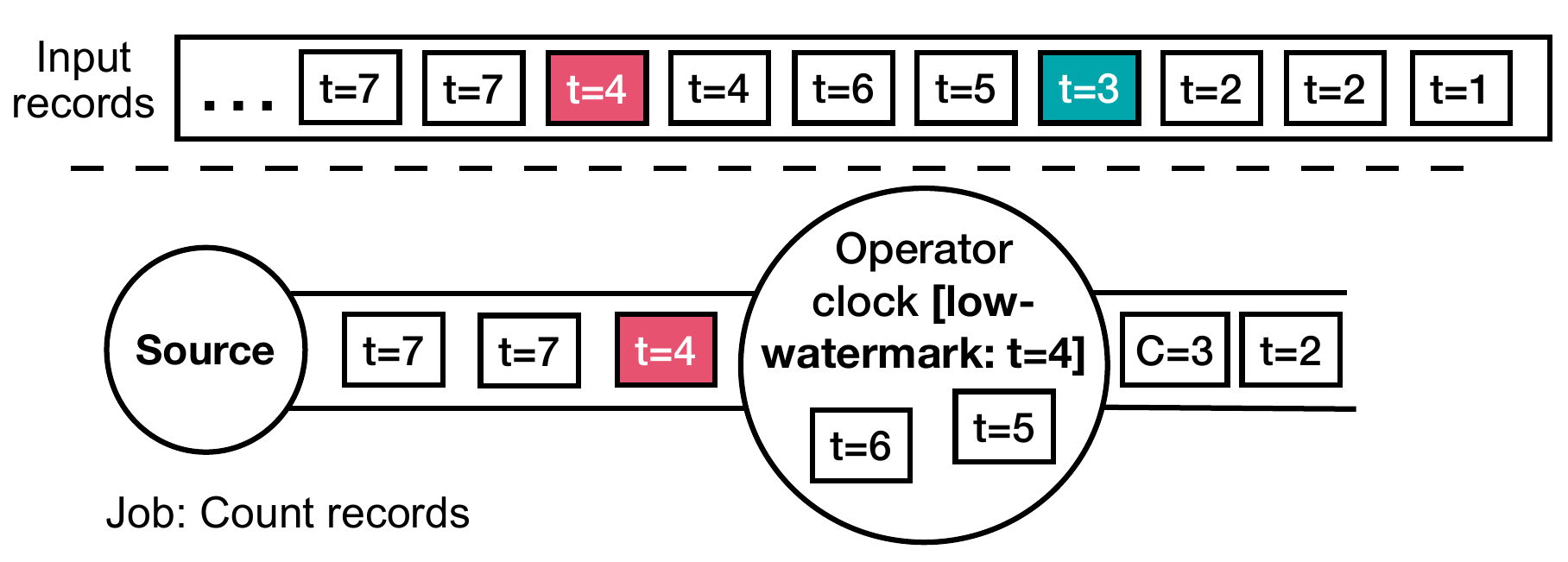}
        \caption{Low-watermark}
        \label{fig:low-watermark}
    \end{subfigure}
    \caption{Mechanisms for managing disorder.}
    \label{fig:disorder-mechanisms}
\end{figure}

We present the four most notable progress tracking mechanisms: slack, heartbeats, low-watermark, and pointstamps.
In addition, we accompany the analysis of each mechanism with a figure. Figure~\ref{fig:disorder-mechanisms} showcases the differences between slack, heartbeats, and low-watermarks.
The figure depicts a simple aggregation operator that counts tuples in 4-second event time tumbling windows.
The operator awaits for some indication that event time has advanced past the end timestamp of a window so that it computes and outputs an aggregate per window.
The indication varies according to the progress-tracking mechanism.
The input to this operator are seven tuples containing only a timestamp from t=1 to t=7.
The timestamp signifies the event time in seconds that the tuple was produced in the input source.
Each tuple contains a different timestamp and all tuples are dispatched from a source in ascending order of timestamp.
Due to network latency, the tuples may arrive to the streaming system out of order.

\paragraph{Slack} is a simple mechanism that involves waiting for out-of-order data for a fixed amount of a certain metric.
Slack originally denoted the number of tuples intervening between the actual occurrence of an out-of-order tuple and the position it would have in the input stream if it arrived on time.
However, it can also be quantified in terms of elapsed time.
Essentially, slack marks a fixed grace period for late tuples.

Figure~\ref{fig:slack} presents the slack mechanism.
In order to accommodate out-of-order tuples the operator admits out-of-order tuples up to \textit{slack=1}.
Thus, the operator having admitted tuples with t=1 and t=2 not depicted in the figure will receive tuple with t=4. 
The timestamp of the tuple coincides with the max timestamp of the first window for interval [0, 4).
Normally, this tuple would cause the operator to  close the window and compute and output the aggregate, but because of the slack value the operator will wait to receive one more tuple.
The next tuple t=3 belongs to the first window and is included there.
At this point, slack also expires and this event finally triggers the window computation, which outputs C=3 for t=[1, 2, 3].
On the contrary, the operator will not accept t=5 at the tail of input because it arrives two tuples after its natural order and is not covered by the slack value.

\paragraph{A heartbeat} is an alternative to slack that consists of an external signal carrying progress information about a data stream.
It contains a timestamp indicating that all succeeding stream tuples will have a timestamp larger than the heartbeat's timestamp.
Heartbeats can either be generated by an input source or deduced by the system by observing environment parameters, such as network latency bound, application clock skew between input sources, and out-of-order data generation~\cite{srivastava2004flexible}.

Figure~\ref{fig:heartbeat} depicts the heartbeat mechanism.
An input manager buffers and orders the incoming tuples by timestamp.
The number of tuples buffered, two in this example (t=5, t=6), is of no importance.
The source periodically sends a heartbeat to the input manager, i.e. a signal with a timestamp.
Then the input manager dispatches to the operator all tuples with timestamp less or equal to the timestamp of the heartbeat in ascending order.
For instance, when the heartbeat with timestamp t=2 arrives in the input manager (not shown in the figure), the input manager dispatches the tuples with timestamp t=1 and t=2 to the operator.
The input manager then receives tuples with t=4, t=6, and t=5 in this order and puts them in the right order.
When the heartbeat with timestamp t=4 arrives, the input manager dispatches the tuple with timestamp t=4 to the operator.
This tuple triggers the computation of the first window for interval [0, 4).
The operator outputs C=2 counting two tuples with t=[1, 2] not depicted in the figure.
The input manager ignores the incoming tuple with timestamp t=3 as it is older than the latest heartbeat with timestamp t=4.

\paragraph{The low-watermark} for an attribute \textit{A} of a stream is the lowest value of \textit{A} within a certain subset of the stream.
Thus, future tuples will probabilistically bear a higher value than the current low-watermark for the same attribute.
Often, \textit{A} is a tuple's event time timestamp.
The mechanism is used by a streaming system to track processing progress via the low-watermark for \textit{A}, to admit out-of-order data whose attribute \textit{A}'s value is not smaller than the low-watermark. Further, it can be used to remove state that is maintained for \textit{A}, such as the corresponding hash table entries of a streaming join computation.

Figure~\ref{fig:low-watermark} presents the low-watermark mechanism, which signifies the oldest pending work in the system.
Here punctuations carrying the low-watermark timestamp decide when windows will be closed and computed.
After receiving two tuples with t=1 and t=2, the corresponding low-watermark for t=2 (which is propagated downstream), and tuple t=3, the operator receives tuple t=5.
Since this tuple carries an event time timestamp greater or equal to 4, which is the end timestamp of the first window, it could be the one to cause the window to fire or close.
However, this approach would not account for out-of-order data.
Instead, the window closes when the operator receives the low-watermark with t=4.
At this point, the operator computes C=3 for t=[1, 2, 3] and assigns tuples with t=[5, 6] to the second window with interval [4, 8).
The operator will not admit tuple t=4 because it is not greater (more recent) than the current low-watermark value t=4.

\paragraph{Comparison between heartbeats, slack, and punctuations.} Heartbeats and slack are both external to a data stream.
Heartbeats are signals communicated from an input source to a streaming system's ingestion point.
Differently to heartbeats, which is an internal mechanism of a streaming system hidden from users, slack is part of the query specification provided by users~\cite{abadi2003aurora}.

Heartbeats and low-watermarks are similar in terms of 
progress-tracking logic.
However, two important differences set them apart.
While heartbeats expose the progress of stream tuple generation at the input sources, the low-watermark extends this to the processing progress of computations within the streaming system by reflecting their oldest pending work.
Second, the low-watermark generalizes the concept of the oldest value, which signifies the current progress point, to any progressing attribute of a stream tuple besides timestamps.

In contrast to heartbeats and slack, \textit{punctuations} are metadata annotations embedded in data streams.
A punctuation is itself a stream tuple, which consists of a set of patterns each identifying an attribute of a stream data tuple.
A punctuation is a generic mechanism that communicates information across the dataflow graph.
Regarding progress tracking, it provides a channel for communicating progress information such as a tuple attribute's low-watermark produced by an operator~\cite{li2008out}, event time skew~\cite{srivastava2004flexible}, or slack~\cite{abadi2003aurora}.
Thus, punctuations can convey which data cease to appear in an input stream; for instance the data tuples with smaller timestamp than a specific value.
Punctuations are useful in other functional areas of a streaming system as well, such as state management, monitoring, and flow control.

\begin{figure*}[t]
    \centering
    \begin{subfigure}{.4\linewidth}
        \centering
        \includegraphics[width=.9\linewidth]{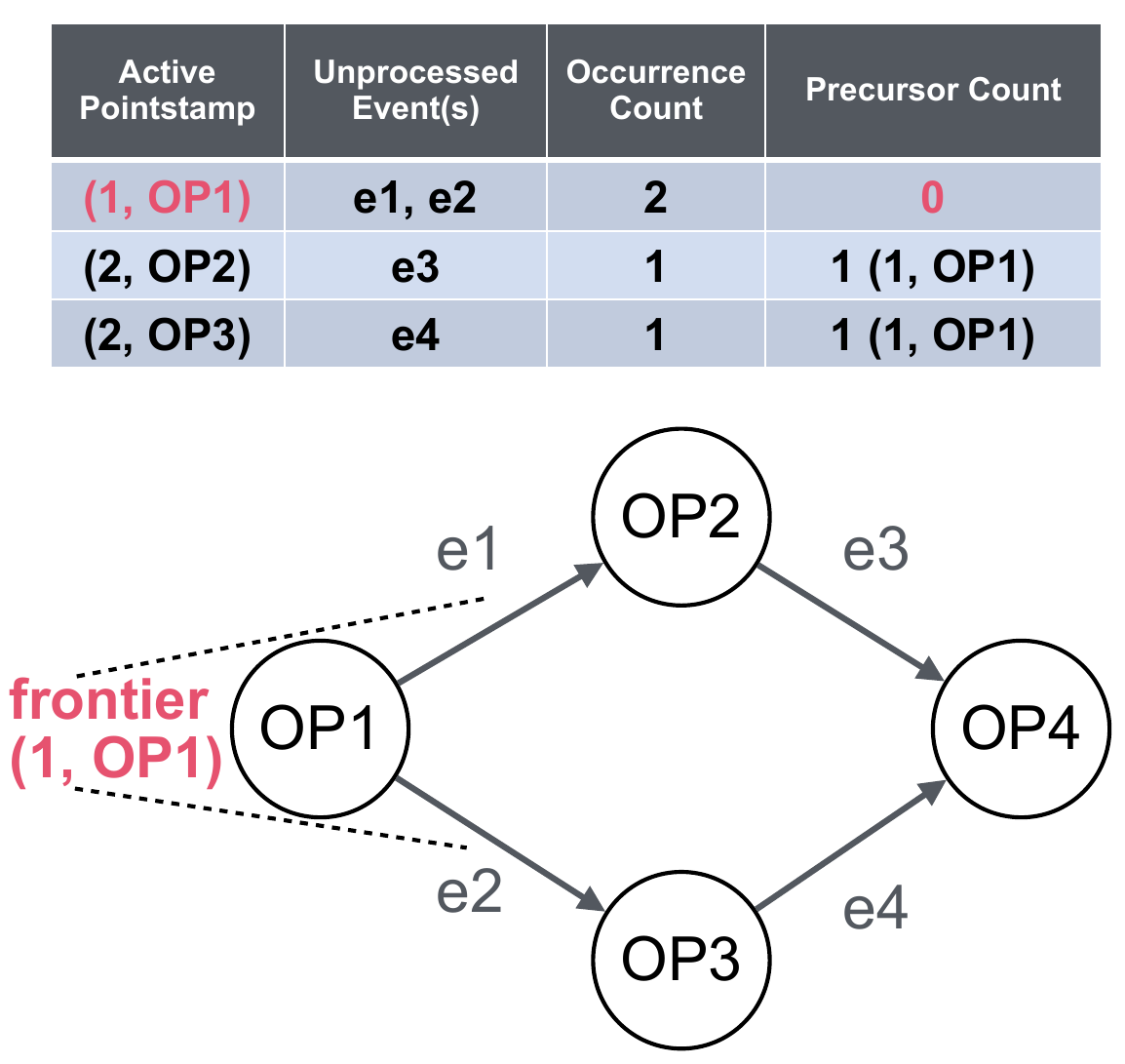}
        \caption{Pointstamps and frontier}
        \label{fig:pointstamp-1}
    \end{subfigure}
    \begin{subfigure}{.4\linewidth}
        \centering
        \includegraphics[width=.8\linewidth]{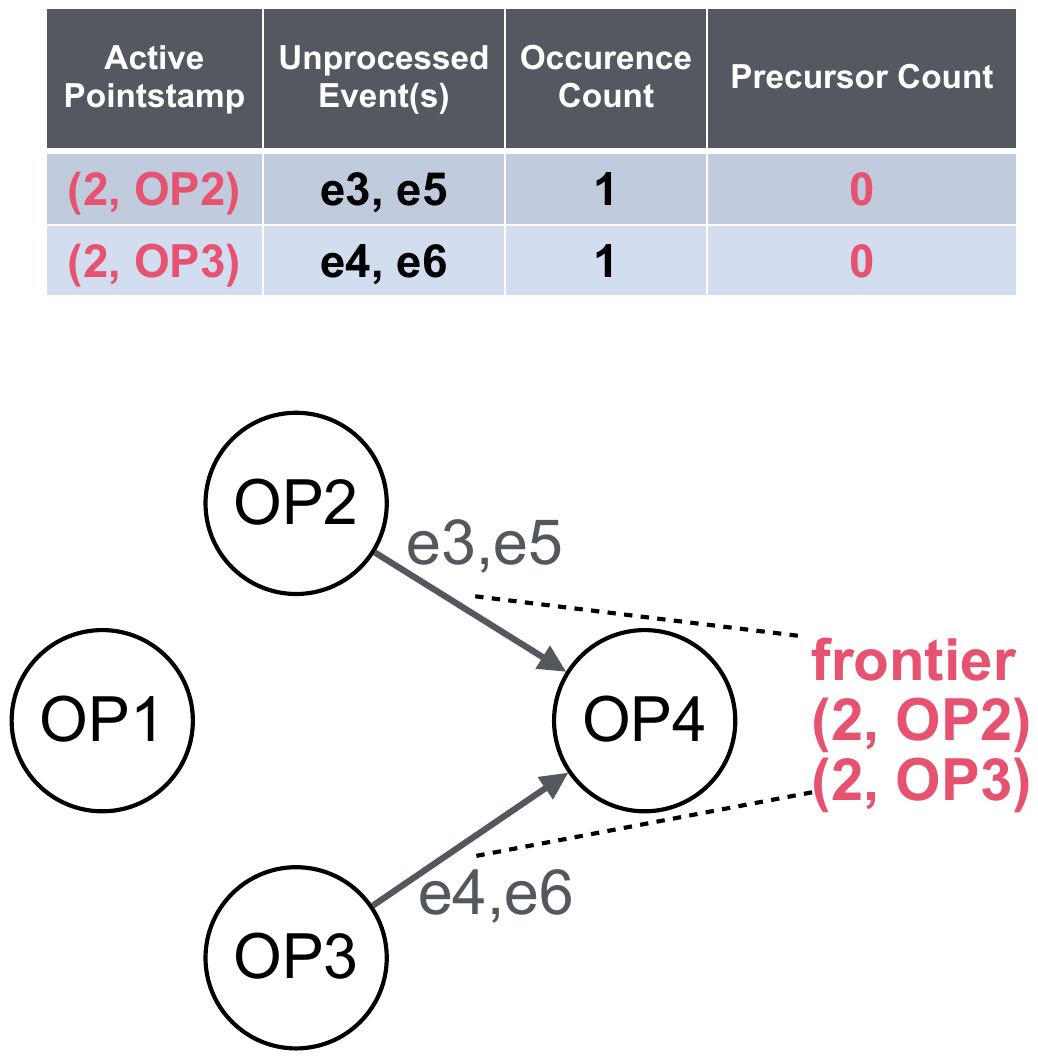}
        \caption{Frontier moves forward}
        \label{fig:pointstamp-2}
    \end{subfigure}
    \caption{High-level workflow of pointstamps and frontier}
    \label{fig:pointstamps}
\end{figure*}

\paragraph{Pointstamps,} like punctuations, are embedded in data streams, but a pointstamp is attached to each stream data tuple as opposed to a punctuation, which forms a separate tuple.
Pointstamps are pairs of timestamp and location that position data tuples on a vertex or edge of the dataflow graph at a specific point in time.
An unprocessed tuple \textit{p} at a specific location \textit{could-result-in} another unprocessed tuple \textit{p'} with timestamp \textit{t'} at another location when \textit{p} can arrive at \textit{p'} before or at timestamp \textit{t'}.
Unprocessed tuples p with timestamp t are in the frontier of processing progress when no other unprocessed tuples could-result-in p.
Thus, tuples bearing \textit{t} or an earlier timestamp are processed and the frontier moves on.
The system enforces that future tuples will bear a greater timestamp than the tuples that generated them.
This modeling of processing progress traces the course of data tuples on the dataflow graph with timestamps and tracks the dependencies between unprocessed events in order to compute the current frontier.
The concept of a frontier is similar to a low-water mark.

The example shown in Figure~\ref{fig:pointstamps} showcases how pointstamps and frontiers work.
The example in Figure~\ref{fig:pointstamp-1} includes three active pointstamps.
Poinstamps are active when they correspond to one or more unprocessed events.
Pointstamp (1, OP1) is in the frontier of active pointstamps, because its precursor count is 0.
The precursor count, specifies the number of active pointstamps that could-result-in that pointstamp.
In the frontier, notifications for unprocessed events can be delivered.
Thus, unprocessed events e1 and e2 can be delivered to OP2 and OP3 respectively.
The occurrence count is 2 because both events e1 and e2 bear the same pointstamp.
Looking at this snapshot of the data flow graph, it is easy to see that pointstamp (1, OP1)
could-result-in pointstamps (2, OP2) and (2, OP3). Therefore, the precursor count of the latter two pointstamps is 1.
A bit later as Figure~\ref{fig:pointstamp-2} depicts, after events e1 and e2 are delivered to OP2 and OP3 respectively, their processing results in the generation of
new events e5 and e6, which bear the same pointstamp as unprocessed events e3 and e4 respectively.
Since there are no more unprocessed events with timestamp 1, and the precursor count of pointstamps (2, OP2) and (2, OP3) is 0,
then the frontier moves on to these active pointstamps.
Consequently, all four event notifications can be delivered.
Obsolete pointstamps (1, OP1), (2, OP2), and (2, OP3), are removed from their location, since they correspond to no unprocessed events.
Although this example is made simple for educational purposes, the progress tracking mechanism, has the power
to track the progress of arbitrary iterative and nested computations.

Pointstamps/frontiers track processing progress regardless of the notion of event time.
However, it is possible for users to capture out-of-order data with pointstamps/frontiers by establishing a two-dimensional frontier of event time and processing time that is flexibly open on the side of event time.

\subsubsection{Tracking progress of out-of-order data in cyclic queries}
Cyclic queries require special treatment for tracking progress.
A cyclic query always contains a binary operator, such as a join or a union.
The output produced by the binary operator meets a loop further in the dataflow graph that connects back to one of the binary operator's input channels.
In a progress model that uses punctuations for instance, the binary operator forwards a punctuation only when it appears in both of its input channels otherwise it blocks waiting for both to arrive.
Since one of the binary operator's input channels depends on its own output channel, a deadlock is inevitable.

Chandramouli et al.~\cite{chandramouli2009fly} propose an operator for detecting progress in cyclic streaming queries on the fly.
The operator introduces a speculative punctuation in the loop that is derived from the passing events' timestamp.
While the punctuation flows in the loop the operator observes the stream's tuples to validate its guess.
When this happens and the speculative punctuation re-enters the operator, it becomes a regular punctuation that carries progress information downstream.
Then a new speculative punctuation is generated and is fed in the loop.
By combining a dedicated operator, speculative output, and punctuations this work achieves to track progress and tolerate disorder in cyclic streaming queries.
The approach works for strongly convergent queries and can be utilized in systems that provide speculative output.

In Naiad~\cite{murray2013naiad, murray2016incremental}, the general progress-tracking model features logical multidimensional timestamps attached to events.
Each timestamp consists of the input batch to which an event belongs and an iteration counter for each loop the event traverses.
Like in Chandramouli et al.~\cite{chandramouli2009fly}, Naiad supports cyclic queries by utilizing a special operator.
However, the operator is used to increment the iteration counter of events entering a loop.
To ensure progress, the system allows event handlers to dispatch only messages with larger timestamp than the timestamp of events being currently processed.
This restriction imposes a partial order over all pending events.
The order is used to compute the earliest logical time of events' processing completion in order to deliver notifications for producing output. Naiad's progress-tracking mechanism is external to the dataflow.
This design defies the associated implementation complexity in favor of a) efficient delivery of notifications that is proportional to dataflow nodes instead of edges and b) incremental computation that avoids redundant work.
Although not directly incorporated, the notion of event time can be encapsulated in multidimensional timestamps to account for out-of-order data.

\subsubsection{Revision processing}
\label{subsubsec:revision-processing}
Revision processing is the update of computations in face of late, updated, or retracted data, which require the modification of previous outputs in order to provide correct results.
Revision processing made its debut in Borealis~\cite{abadi2005design}.
From there on, it has been combined with in-order processing architectures~\cite{Mutschler2013Reliable,Chandramouli2018Impatience}, as well as
out-of-order processing architectures~\cite{Barga2007Consistent, Krishnamurthy2010Continuous, Ali2011StreamInsight, akidau2015dataflow}.
In some approaches revision processing works by \textit{storing} incoming data and \textit{revising} computations in face of late, updated, or retracted data~\cite{Barga2007Consistent, Ali2011StreamInsight, akidau2015dataflow}.
Other approaches \textit{replay} affected data, \textit{revise} computations, and propagate the revision messages to update all affected results until the present~\cite{abadi2005design, Ryvkina2006Revision, Mutschler2013Reliable}.
Finally, a third line of approaches maintain multiple \textit{partitions} that capture events with different levels of lateness and \textit{consolidate} partial results~\cite{Krishnamurthy2010Continuous, Chandramouli2018Impatience}.


\paragraph{Store and revise.} Microsoft's CEDR~\cite{Barga2007Consistent} and StreamInsight~\cite{Ali2011StreamInsight}, and Google's Dataflow~\cite{akidau2015dataflow} buffer or store stream data and process late events, updates, and deletions incrementally by revising the captured values and updating the computations.

The dataflow model~\cite{akidau2015dataflow} divides the concerns for out-of-order data into three dimensions: the event time when late data are processed, the processing time when corresponding results are materialized, and how later updates relate to earlier results.
The mechanism that decides the emission of updated results and how the refinement will happen is called a \textit{trigger}.
Triggers are signals that cause a computation to be repeated or updated when a set of specified rules fire.

One important rule regards the arrival of late input data.
Triggers ensure output correctness by incorporating the effects of late input into the computation results.
Triggers can be defined based on watermarks, processing time, data arrival metrics,
and combinations of those; they can also be user-defined.
Triggers support three refinement policies, accumulating where new results overwrite older ones, discarding where new results complement older ones, and accumulating and retracting where new results overwrite older ones and older results are retracted.
Retractions, or compensations, are also supported in StreamInsight~\cite{Ali2011StreamInsight}.


\paragraph{Replay and revise.} \emph{Dynamic revision}~\cite{abadi2005design} and \emph{speculative processing}~\cite{Mutschler2013Reliable} replay an affected past data subset when a revision tuple is received.
An optimization of this scheme relies on two revision processing mechanisms, upstream processing and downstream processing~\cite{Ryvkina2006Revision}.
Both are based on a special-purpose operator, called \emph{connection point}, that intervenes between two regular operators and stores tuples output by the upstream operator.
According to the upstream revision processing, an operator downstream from a connection point can ask for a set of tuples to be replayed so that it can calculate revisions based on old and new results.
Alternatively, the operator can ask from the downstream connection point to retrieve a set of output tuples related to a received revision tuple.
Under circumstances, the operator can calculate correct revisions by incorporating the net effect of the difference between the original tuple and its revised one to the old result.



Dynamic revision emits delta revision messages, which contain the difference of the output between the original and the revised value.
It keeps the input message history to an operator in the connection point of its input queue. Since keeping all messages is infeasible, there is a bound in the history of messages kept.
Messages that go further back from this bound can not be replayed and, thus, revised.
Dynamic revision differentiates between stateless and stateful operators.
A stateless operator will evaluate both the original $(t)$ and the revised message $(t')$ emitting the delta of their output.
For instance, if the operator is a filter, $t$ is true and $t'$ is not, then the operator will emit a deletion message for $t$.
A stateful operator, on the other hand, has to process many messages in order to emit an output. Thus, an aggregation operator has to re-process the whole window for both a revised message and the original message contained in that window in order to emit revision messages.
Dynamic revision is implemented in Borealis.


Speculative processing, on the other hand, applies snapshot recovery if no output has been produced for a disordered input stream.
Otherwise, it retracts all produced output in a  recursive manner.
In speculative processing because revision processing is opportunistic, no history bound is set.






\setlength{\tabcolsep}{5pt}
\begin{table*}
\smaller\centering
\caption{Event order management in streaming systems}
\label{tab:disorder-management-streaming}
\begin{tabular}{p{0.09\textwidth}
                        p{0.05\textwidth}
                        p{0.07\textwidth}
                        p{0.05\textwidth}
                        p{0.1\textwidth}
                        p{0.1\textwidth}
                        p{0.2\textwidth}
                        p{0.18\textwidth}
                        }
\hline
\multicolumn{1}{c}{\textbf{System}} &
\multicolumn{3}{c}{\textbf{Architecture}} &
\multicolumn{4}{c}{\textbf{Progress-tracking}} \\
\rowcolor{white}
& \multicolumn{1}{c}{In-order}
& \multicolumn{1}{c}{Out-of-order}
& \multicolumn{1}{c}{Revision}
& \multicolumn{1}{c}{Mechanism}
& \multicolumn{1}{c}{Communication}
& \multicolumn{1}{c}{Disorder bound metric}
& \multicolumn{1}{c}{Revision approach} \\
\hline

Aurora*~\cite{cherniack2003scalable, abadi2003aurora}
& \multicolumn{1}{a}{\checkmark}
& \multicolumn{1}{a}{}
& \multicolumn{1}{a}{}
& \multicolumn{1}{d}{Slack}
& \multicolumn{1}{d}{User config}
& \multicolumn{1}{d}{Number of tuples}
& \multicolumn{1}{d}{---}
\\
\hline

STREAMS~\cite{srivastava2004flexible}
& \multicolumn{1}{a}{\checkmark}
& \multicolumn{1}{a}{}
& \multicolumn{1}{a}{}
& \multicolumn{1}{d}{Heartbeat}
& \multicolumn{1}{d}{Signal to input manager}
& \multicolumn{1}{d}{Timestamp (event time skew, net-}
& \multicolumn{1}{d}{---}
\\

& \multicolumn{1}{a}{}
& \multicolumn{1}{a}{}
& \multicolumn{1}{a}{}
& \multicolumn{1}{d}{}
& \multicolumn{1}{d}{}
& \multicolumn{1}{d}{work latency, out-of-order bound)}
& \multicolumn{1}{d}{}
\\
\hline

Borealis~\cite{abadi2005design}
& \multicolumn{1}{a}{}
& \multicolumn{1}{a}{\checkmark}
& \multicolumn{1}{a}{\checkmark}
& \multicolumn{1}{d}{History bound}
& \multicolumn{1}{d}{System config}
& \multicolumn{1}{d}{Number of tuples or time units}
& \multicolumn{1}{d}{Replay past data, enter revised}
\\

& \multicolumn{1}{a}{}
& \multicolumn{1}{a}{}
& \multicolumn{1}{a}{}
& \multicolumn{1}{d}{}
& \multicolumn{1}{d}{}
& \multicolumn{1}{d}{}
& \multicolumn{1}{d}{values, issue delta output}
\\
\hline

Gigascope~\cite{Johnson2005A}
& \multicolumn{1}{a}{}
& \multicolumn{1}{a}{\checkmark}
& \multicolumn{1}{a}{}
& \multicolumn{1}{d}{Low-watermark}
& \multicolumn{1}{d}{Punctuation}
& \multicolumn{1}{d}{Timestamp}
& \multicolumn{1}{d}{---}
\\
\hline

Timestream~\cite{QianHS13}
& \multicolumn{1}{a}{\checkmark}
& \multicolumn{1}{a}{}
& \multicolumn{1}{a}{}
& \multicolumn{1}{d}{Low-watermark}
& \multicolumn{1}{d}{Punctuation}
& \multicolumn{1}{d}{Timestamp}
& \multicolumn{1}{d}{---}
\\
\hline

Millwheel~\cite{akidau2013millwheel}
& \multicolumn{1}{a}{}
& \multicolumn{1}{a}{\checkmark}
& \multicolumn{1}{a}{}
& \multicolumn{1}{d}{Low-watermark}
& \multicolumn{1}{d}{Signal to central authority}
& \multicolumn{1}{d}{Timestamp}
& \multicolumn{1}{d}{---}
\\
\hline

Naiad~\cite{murray2013naiad}
& \multicolumn{1}{a}{}
& \multicolumn{1}{a}{\checkmark}
& \multicolumn{1}{a}{\checkmark}
& \multicolumn{1}{d}{Pointstamp}
& \multicolumn{1}{d}{Part of data tuple}
& \multicolumn{1}{d}{Multidimensional timestamp}
& \multicolumn{1}{d}{Incremental processing of up-}
\\

& \multicolumn{1}{a}{}
& \multicolumn{1}{a}{}
& \multicolumn{1}{a}{}
& \multicolumn{1}{d}{}
& \multicolumn{1}{d}{}
& \multicolumn{1}{d}{}
& \multicolumn{1}{d}{dated data via structured loops}
\\
\hline

Trill~\cite{chandramouli2014trill}
& \multicolumn{1}{a}{\checkmark}
& \multicolumn{1}{a}{}
& \multicolumn{1}{a}{}
& \multicolumn{1}{d}{Low-watermark}
& \multicolumn{1}{d}{Punctuation}
& \multicolumn{1}{d}{Timestamp}
& \multicolumn{1}{d}{---}
\\
\hline

Streamscope~\cite{LinHZ16}
& \multicolumn{1}{a}{\checkmark}
& \multicolumn{1}{a}{}
& \multicolumn{1}{a}{}
& \multicolumn{1}{d}{Low-watermark}
& \multicolumn{1}{d}{Punctuation}
& \multicolumn{1}{d}{Timestamp; sequence number}
& \multicolumn{1}{d}{---}
\\
\hline

Samza~\cite{NoghabiPP17}
& \multicolumn{1}{a}{}
& \multicolumn{1}{a}{\checkmark}
& \multicolumn{1}{a}{\checkmark}
& \multicolumn{1}{d}{Low-watermark}
& \multicolumn{1}{d}{Punctuation}
& \multicolumn{1}{d}{Timestamp}
& \multicolumn{1}{d}{Find, roll back, recompute af-}
\\

& \multicolumn{1}{a}{}
& \multicolumn{1}{a}{}
& \multicolumn{1}{a}{}
& \multicolumn{1}{d}{}
& \multicolumn{1}{d}{}
& \multicolumn{1}{d}{}
& \multicolumn{1}{d}{fected input windows}
\\
\hline

Flink~\cite{carbone2015apache}
& \multicolumn{1}{a}{}
& \multicolumn{1}{a}{\checkmark}
& \multicolumn{1}{a}{\checkmark}
& \multicolumn{1}{d}{Low-watermark}
& \multicolumn{1}{d}{Punctuation}
& \multicolumn{1}{d}{Timestamp}
& \multicolumn{1}{d}{Store \& Recompute/Revise}
\\
\hline

Dataflow~\cite{akidau2015dataflow}
& \multicolumn{1}{a}{}
& \multicolumn{1}{a}{\checkmark}
& \multicolumn{1}{a}{\checkmark}
& \multicolumn{1}{d}{Low-watermark}
& \multicolumn{1}{d}{Signal to central authority}
& \multicolumn{1}{d}{Timestamp}
& \multicolumn{1}{d}{Discard and recompute; accu-}
\\

& \multicolumn{1}{a}{}
& \multicolumn{1}{a}{}
& \multicolumn{1}{a}{}
& \multicolumn{1}{d}{}
& \multicolumn{1}{d}{}
& \multicolumn{1}{d}{}
& \multicolumn{1}{d}{mulate and revise; custom}
\\
\hline

Spark~\cite{armbrust2018structured}
&  \multicolumn{1}{a}{}
& \multicolumn{1}{a}{\checkmark}
& \multicolumn{1}{a}{\checkmark}
& \multicolumn{1}{d}{Slack}
& \multicolumn{1}{d}{User config}
& \multicolumn{1}{d}{Number of seconds}
& \multicolumn{1}{d}{Discard and recompute; accu-}
\\

& \multicolumn{1}{a}{}
& \multicolumn{1}{a}{}
& \multicolumn{1}{a}{}
& \multicolumn{1}{d}{}
& \multicolumn{1}{d}{}
& \multicolumn{1}{d}{}
& \multicolumn{1}{d}{mulate and revise}
\\
\hline

\end{tabular}
\end{table*}

\paragraph{Partition and consolidate.}
Both \emph{order-independent processing}~\cite{Krishnamurthy2010Continuous} and \emph{impatience sort}~\cite{Chandramouli2018Impatience} are based on partial processing of independent partitions in parallel and consolidation of partial results.
In order-independent processing, when a tuple is received after its corresponding progress indicator a new partition is opened and a new query plan instance processes this partition using standard out-of-order processing techniques.
On the contrary, in impatience sort, the latest episode of the vision of CEDR~\cite{Barga2007Consistent}, an online sorting operator incrementally orders the input arriving at each partition so that it is emitted in order.
The approach uses punctuations to bound the disorder as opposed to order-independent processing which can handle events arriving arbitrarily late.

In order-independent processing, partitioning is left for the system to decide while in impatience sort it is specified by the users.
In order-independent processing, tuples that are too old to be considered in their original partition are included in the partition which has the tuple with the closest data.
When no new data enter an ad-hoc partition for a long time, the partition is closed and destroyed by means of a heartbeat.
Ad-hoc partitions are window-based; when an out-of-order tuple is received that does not belong to one of the ad-hoc partitions, a new ad-hoc partition is introduced.
An out-of order tuple with a more recent timestamp than the window of an ad-hoc partition causes that partition to flush results and close.
Order-independent processing is implemented in Truviso.

On the contrary, in impatience sort, users specify reorder latencies, such as $1ms$, $100ms$, and $1s$, that define the buffering time for ingesting and sorting out-of-order input tuples.
According to the specified reorder latencies, the system creates different partitions of in-order input streams.
After sorting, a union operator merges and synchronizes the output of a partition $P$ with the output of a partition $L$ that features lower reorder latency than $P$.
Thus, the output will incorporate partial results provided by $L$ with later updates that $P$ contains.
This way applications that require fast but partial results can subscribe to a partition with small reorder latency and vice versa.
By letting applications choose the desired extent of reorder latency this design provides for different trade-offs between completeness and freshness of results.
Impatience sort is implemented in Microsoft Trill.

\subsection{1st generation vs. 2nd generation}
\label{sub:vintage}


The importance of event order in data stream processing became obvious since its early days~\cite{babcock2002models} leading to the first wave of simple intuitive solutions. Early approaches involved buffering and reordering arriving tuples using some measure for adjusting the frequency and lateness of data dispatched to a streaming system in order~\cite{chandrasekaran2003telegraphcq, abadi2003aurora, srivastava2004flexible}.
A few years later, the introduction of out-of-order processing~\cite{li2008out} improved throughput, latency, and scalability for window operations by keeping track of processing progress without ordering tuples.
In the meantime, revision processing~\cite{abadi2005design} was proposed as a strategy for dealing with out-of-order data reactively.
In the years to come, in-order, out-of-order, and revision processing were extensively explored, often in combination with one another~\cite{Barga2007Consistent, Krishnamurthy2010Continuous, Ali2011StreamInsight, Mutschler2013Reliable, akidau2015dataflow}.
Modern streaming systems implement a refinement of these original concepts.
Interestingly, concepts devised several years ago, like low-watermarks, punctuations, and triggers, which advance the original revision processing, were popularized recently by streaming systems such as Millwheel~\cite{akidau2013millwheel} and the Google Dataflow model~\cite{akidau2015dataflow}, Flink~\cite{carbone2015apache}, and Spark~\cite{armbrust2018structured}. Table~\ref{tab:disorder-management-streaming} presents how both 1st generation and modern streaming systems implement out-of-order data management.

\subsection{Open Problems}
\label{sub:open}

Managing data disorder entails architecture support and flexible mechanisms.
There are open problems at both levels.


First, which architecture is better is an open debate. Although many of the latest streaming systems adopt an out-of-order architecture, opponents finger the architecture's implementation and maintainance complexity. In addition, revision processing, which is used to reconcile out-of-order tuples is daunting at scale because of the challenging state size. On the other hand, in-order processing is resource-hungry and loses events if they arrive after the disorder bound.
    
Second, applications receiving data streams from different sources may need to support multiple notions of event time, one per incoming stream, for instance.
However, streaming systems to date cannot support multiple time domains.

Finally, data streams from different sources may have disparate latency characteristics that render their watermarks unaligned.
Tracking the processing progress of those applications is challenging for today's streaming systems.




\section{State Management}\label{sec:state}

State is effectively what captures all internal side-effects of a continuous stream computation. This includes, for example, active windows, buckets of records, partial or incremental aggregates used in an application as well as possibly some user-defined variables created and updated during the execution of a stream pipeline. A careful look into how state is exposed and managed in stream processing systems unveils an interesting trace of trends in computer systems and cloud computing as well as a revelation of prospects on upcoming capabilities in event-based computing. This section provides an overview of known approaches, modern directions and discussions of open problems in the context of state management.

\setlength{\tabcolsep}{5pt}
\begin{table*}
\smaller\centering
\caption{State Management Features in Streaming Systems}
\label{tab:state}
{
\begin{tabular}{c|a|d|a|a|a|d} 
\hline
\begin{tabular}[c]{@{}l@{}}\\\textbf{System}\end{tabular} 
& \multicolumn{1}{c}{\textbf{Programmable State}} & 
\multicolumn{1}{c}{\textbf{State Mgmt Responsibility}} & 
\multicolumn{3}{c}{\textbf{State Mgmt Architecture}}                              & \multicolumn{1}{c}{\textbf{Persistence Granularity}}  \\
                                                          &                        \multicolumn{1}{c}{}                         &   \multicolumn{1}{c}{}                                                     & \multicolumn{1}{c}{In-memory}                 & \multicolumn{1}{c}{Out-of-Core}               & \multicolumn{1}{c}{External}                  &     \multicolumn{1}{c}{}                                                  \\ 
\hline
Aurora/Borealis~\cite{cherniack2003scalable}                                          & \xmark                           & System                                                 & \checkmark &                           &                           & No persistence                                        \\ 
\hline
STREAM~\cite{arasu2004stream}                                                   & \xmark                           & System                                                 & \checkmark &                           &                           & No persistence                                        \\ 
\hline
TelegraphCQ~\cite{ShahHB04}                                              & \xmark                           & System                                                 & \checkmark &                           &                           & No persistence                                        \\ 
\hline
S4~\cite{neumeyer2010s4}                                                      & \checkmark                       & User                                                   &                           &                           & \checkmark & No persistence                                        \\ 
\hline
Storm (1.0)~\cite{toshniwal2014storm}                                              & \checkmark                       & User                                                   &                           &                           & \checkmark & No Persistence                                        \\ 
\hline
Spark(1.0)~\cite{armbrust2018structured}                                               & \checkmark                       & System                                                 & \checkmark &                           &                           & Batch-level                                           \\ 
\hline
Trident~\cite{CUSTOM:web/trident}                                                  & \checkmark                       & System                                                 & \checkmark & \checkmark &                           & Batch-level                                           \\ 
\hline
SEEP~\cite{fernandez2014making}                                                     & \checkmark                       & System                                                 & \checkmark &                           &                           & Epoch-level                                           \\ 
\hline
Naiad~\cite{murray2013naiad}                                                    & \checkmark                       & System                                                 & \checkmark &                           &                           & Epoch-level                                           \\ 
\hline
TimeStream~\cite{QianHS13}                                               & \checkmark                       & System                                                 & \checkmark &                           &                           & Epoch-level                                           \\ 
\hline
Millwheel~\cite{akidau2013millwheel}                                                & \checkmark                       & System                                                 &                           &                           & \checkmark & Record-level                                          \\ 
\hline
Flink~\cite{CarboneEF17, carbone2015apache}                                                    & \checkmark                       & System                                                 & \checkmark & \checkmark &                           & Epoch-level                                           \\ 
\hline
Kafka-Streams~\cite{CUSTOM:web/kafkastreams}                                            & \xmark                           & System                                                 & \checkmark & \checkmark &                           & Epoch-level                                           \\ 
\hline
Samza~\cite{NoghabiPP17}                                                    & \checkmark                       & System                                                 & \checkmark & \checkmark &                           & Epoch-level                                           \\ 
\hline
Streamscope~\cite{LinHZ16}                                              & \checkmark                       & System                                                 & \checkmark &                           &                           & Epoch-level                                           \\ 
\hline
S-Store~\cite{CetintemelDK14, TatbulZM15}                                                  & \xmark                           & System                                                 &                           &                           & \checkmark & Batch-level                                           \\ 
\hline                                                     
\end{tabular}
}
\end{table*}

\subsection{Managing Stream Processing State}
Stream state management is still an active research field, incorporating  methods on how state should be declared in a stream application, as well as how it should be scaled and partitioned. Furthermore, state management considers state persistence methods infinite/long running applications and defines system guarantees and properties to maintain whenever a change in the system occurs. 

A change during a system's runtime often requires state reconfiguration. Such a change can be the result of a partial process or network failure, but also actions that need to be taken to adjust compute and storage capacity (e.g., scaling-up/down). Most of these research issues have been introduced in part within the context of pioneering DSMSs such as Aurora and Borealis~\cite{ccetintemel2016aurora}. Specifically, Boralis has set the foundations in formulating many of these problems such as the need for embedded state, persistent store access as well as failure recovery protocols. In \autoref{tab:state} we categorize known data stream processing systems according to their respective state management approaches. The rest of this section offers an overview of each of the topics in stream state management along with past and currently employed approaches, all of which we categorize as follows:

\subsection{Programmability \& Responsibility}
State in a programming model can be either implicitly or explicitly declared and used. We define state programmability as the ability of a streaming system to allow its users to define user-defined state. State in this case can be a local variable within a stateful \texttt{map} function, representing a counter. Programmability in state requires support from the underlying execution engine, a feature that directly affects the engine's complexity. Different system trends have influenced both how state can been exposed in a data stream programming model as well as how it should be scoped and managed. In this section, we discuss different approaches and their trade offs. As shown in \autoref{tab:state}, very few systems disallow their users to define custom user-defined state. These systems focus more on providing a high-level SQL interface on top of a dataflow processor allowing only their internal operators to define and use state within stateful operations (e.g., joins, windows, aggregates).

\stitle{State Management Responsibility} An orthogonal aspect to programmability is state management \textit{responsibility}, which entails the obligation of maintaining state by either the user or the system. State maintenance includes, allocating memory/disk space for storing application variables, persisting changes to disk and recovering state entries from durable storage if needed upon system recovery. The first generation of data parallel stream processing systems, i.e., Storm~\cite{toshniwal2014storm} and S4~\cite{neumeyer2010s4} required user-managed state. In such systems, stateful processing was either implemented with no guarantees making use of custom in-memory data structures or, often implemented using external key-value stores which cover certain scalability and persistence needs. For the rest of the systems available, state management concerns have been internally handled by the streaming systems themselves through the use of explicit state APIs or non-programmable, yet, internally managed state abstractions.

\subsubsection{Discussion}

In the early days of data stream management when main memory was scarce, state had a facilitating role, supporting the implementation of system operators, such as CQL's join, filter, and sort as employed in STREAM~\cite{arasu2004stream}. We term this type of state, defined by the designers of a given system and used by the internal operators of that system, \textit{system-defined state}. A common term used to describe that type of state was ``synopsis''. Typically, users of such systems were oblivious of the underlying state and its implicit nature resembled the use of intermediate results in DBMSs. Systems such as STREAM, as well as Aurora Borealis \cite{ccetintemel2016aurora}, attached special synopses to a stream application's dataflow graph supporting different operators, such as a window max, a join index or input source buffers for offsets. A noteworthy feature in STREAM was the capability to re-use synopses compositionally to define other synopses in an application internally in the system. Overall, synopses have been one of the first forms of state in early stream processing systems primarily for stream processing over shared-memory. Several of the issues regarding state, including fault tolerance and load balancing, were already considered back then, for example in Borealis. Although, the lack of user-defined state limited the expressive power of that generation of systems to a subset of relational operations. Furthermore, the use of over-specialized data structures was somewhat oblivious to the needs of reconfiguration which requires state to be flexible and easy to partition.

In the post-MapReduce era, there was a primary focus in compute scalability with systems like Storm~\cite{CUSTOM:web/Storm} allowing the composition of distributed pipelines of tasks. For application flexibility and simplicity, many of these systems did not provide any state management whatsoever, leaving everything regarding state to the hands of the programmer. That included both declaration and management of state. User-declared and managed state was either defined and used within the working memory and scope provided by the hosting framework or defined and persisted externally, using an existing key value storage or database system (e.g. Redis \cite{CUSTOM:web/redis,leibiusky2012getting}). In summary, application-managed state offers flexibility and gives expert users implementation freedom. However, no state management capabilities are offered from the system's side. As a result, the user has to reason about persistence, out-of-core scalability, and all necessary third-party storage system dependencies. These are all complex choices to make and require a combination of deep expertise and additional engineering work to integrate stream and storage technologies.

Currently, most stream processing systems allow a level of freedom for user-defined state through a form of a stateful processing API. This enriches stream applications to define their custom state, while also granting the underlying system access to state information in order to employ data management mechanisms for persistence, scalability and fault tolerance. State information includes types used, serializers/deserializers and read and write operations known at runtime. The main limitation of user-defined, system-managed state is the lack of direct control on data structures that materialize that state (e.g., for custom optimizations). 

\subsection{State Management Architecture}
\begin{figure}[t!]
        \includegraphics[width=.9\linewidth]{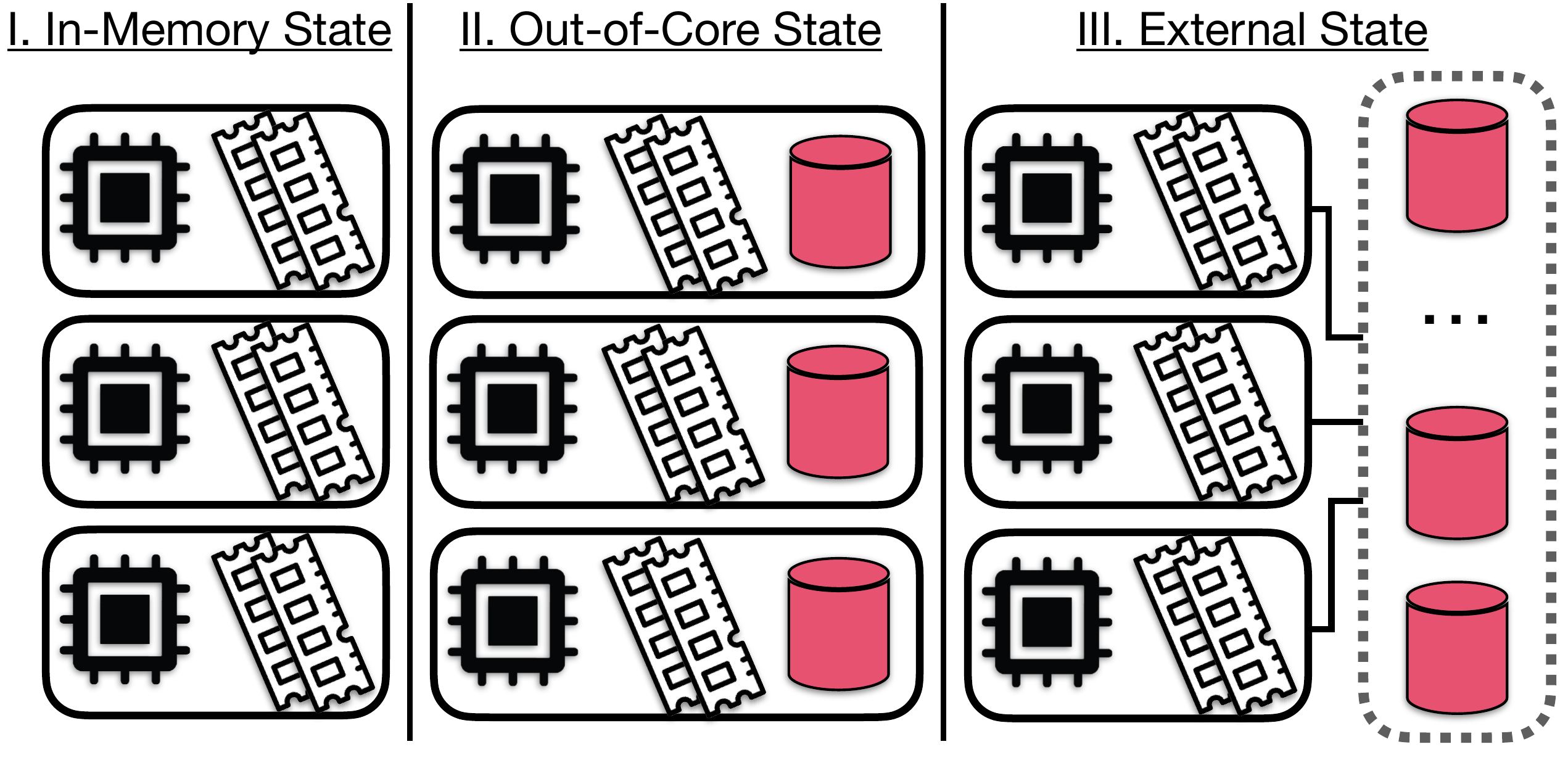}
        \caption{Scalable Architectures for Stateful Data Streaming}
        \label{fig:streamstate}
\end{figure}

The state management architecture refers to the way that a streaming system stores and manages its internal or user-defined state. We observe three distinct stateful processing directions in the architecture of data stream runtime systems (depicted in \autoref{fig:streamstate}): \begin{itemize}
    \item \textbf{In-memory} architectures entail storing active state within in-memory data structures. This approach is able to sustain state bounded within available main-memory available in each node executing stream operators. 
    \item \textbf{Out-of-core} architectures make use of multiple levels of storage mediums such as non-volatile memory to manage state. This approach allows exploiting fast main memory acccess within each compute node while also supporting a growing number of state entries which are split and archived in secondary storage. We observe that the out-of-core data structure of choice used in most systems is a variant of the LSM-Tree \cite{o1996log} such as FASTER~\cite{chandramouli2018faster} or RocksDB/LevelDB\footnote{\url{github.com/google/leveldb}}.
    \item \textbf{External} architectures decouple compute and state, allowing state to be handled by an external database or key-value store. This approach enables more modular system designs (state \& compute decoupling which is very Cloud-friendly) and effective re-use of several desired properties of database systems (e.g., ACID transactions, consistency guarantees, auto-scaling) in support of more complex guarantees in the context of data streaming.  A predominant usage of external state was common within applications in Apache Storm. The lack of system-managed state necessitated users to store all of their state in an external systems. In this architecture, when state access is needed, the streaming operator has to reach out to the external system, dramatically increasing its latency. Google’s Millwheel, the cloud engine of Beam/Google Dataflow is a representative example of system-managed external state architecture. Millwheel builds on the capabilities of BigTable~\cite{chang2008bigtable} and Spanner~\cite{corbett2013spanner} (e.g., blind atomic writes). Tasks in Millwheel are effectively stateless. They do keep recent local changes in memory but overall they commit every single output and state update to BigTable as a single transaction. This means that Millwheel is using an external store for both persisting every single working state per key but also all necessary logs and checkpoints needed for recovery and non-idempontent updates.
    
\end{itemize}

\subsection{Persistence Granularity}
The persistence granularity refers to the granularity in which a streaming system makes a snapshot of the state. While some older systems did not provide any guarantees for the state persistence (e.g., Aurora/Borealis relied on duplicate/standby operators as described in \autoref{sec:FT}). Most systems at the moment employ a coarse-grained persistence. 

Epoch-level persistence granularity is typically achieved in the form of application-level snapshots. Most commonly, systems employ a form of asynchronous consistent snapshotting such as the Chandy-Lamport algorithm \cite{chandy1985distributed} as such: each epoch, i.e., either periodically or after a certain number of records have been ingested by the system, each operator acquires a copy of its state. The batch-level persistence seen in systems such as Spark Streaming, and Trident/Storm adopts a strict micro-batching processing paradigm: i.e., a batch execution is submitted after collecting a sizable number of records, and the state of an operator is stored right after a given batch has been processed. In S-Store the batch granularity orchestrated as a series of ACID transactions on top of a relational database.

Another extreme to the epoch-based approach is record-level persistence. This approach as seen in Millwheel~\cite{akidau2013millwheel} follows a record-level epoch model: it stores the state transition of each operator on every single output (detailed in \autoref{sec:FT}). Section~\ref{sec:stateconsistency} offers an in-depth analysis of the implications between transactional stream processing and persistence granularity.

\subsubsection{Discussion}
Stream processing has been influenced by general trends in scalable computing. State and compute have gradually evolved from a scale-up task-parallel execution model to the more common scale-out data-parallel model with related implications in state representations and operations that can be employed. Persistent data structures have been widely used in database management systems ever since they were conceived. In data stream processing the idea of employing internal and external persistence strategies was uniformly embraced in more recent generations of systems. Section~\ref{sec:statescales} covers different architectures and presents examples of how modern systems can support large volumes of state, beyond what can fit in memory, within unbounded executions. Another foundational transitioning step in stream technology has been the development and adoption of transactional-level guarantees. Section ~\ref{sec:stateconsistency} gives an overview of the state of the art and covers the semantics of transactions in data streaming alongside implementation methodologies.



\subsection{Scalability and State Management}
\label{sec:statescales}

Scalable state has been the main incentive of the second generation of stream processing systems which automated deployment and partitioning of data stream computations. The need for scalable state was driven by the need to facilitate unbounded data stream executions where the space complexity for stream state is linear to the over-increasing input consumed by a stream processor at any point in time. This section discusses types of scalable state, as well as scalable system architectures that can sustain support for partitioning, persisting, and committing changes to large volumes of state.

\subsubsection{Parallel vs. Global Stateful Operations}

Scalable state takes two forms in a stream application, typically referred to as \emph{partitioned} and \emph{non-partitioned} state (also referred to as global state). Depending on the nature of a specific operation, one or both of these state types can be employed.

\stitle{Partitioned State.}
Partitioned state is the de facto way to enable data-parallel computation on massive data streams.  Partitioned state allows key-wise logical partitioning of state to compute tasks, where each logical task handles a specific key. This is enabled in the API level through an additional operation that is invoked prior to stateful processing which lifts the scope from task- to key-based processing such as ``keyBy'' in Apache Flink or ``groupBy'' in Beam and Kafka-Streams. At the physical level, multiple keys (or key ranges) can be assigned to a specific physical task or compute node.

\stitle{Non-partitioned State.}
Non-partitioned state is mapped as a singleton to physical compute tasks. Such non-partitioned state is typically used in two ways. First, in order to compute global aggregates over the complete input stream. Second, it can be used to calculate aggregates at the level of the physical operator (e.g., count how many keys have been processed per operator). Task-level state can also be useful for keeping offsets when consuming logs from  a physical stream source task. Because non-partitioned state either deals with operator-local computations or with global aggregates, its use is not scalable and should be used with caution by practitioners.

\subsection{Managing Consistency and Persistence}
\label{sec:stateconsistency}

Consistent stream processing has for long been an open research issue due to the challenging nature of distributed unbounded processing but also due to the lack of a formal specification of the problem itself. Consistency relates to guarantees a system can make at the face of failure as well as any need for change during its operation. 
In data streaming, changing or updating a running data stream application is a concept also known as reconfiguration. For example, this includes the case when one needs to apply a software update to a stream application or scale out to more compute nodes without loss of accuracy or computation. The underlying relation between fault tolerance and reconfiguration has been highlighted by several works in the past such as the research behind the SEEP system \cite{castro2013integrating} that considers an integrated approach to scale and recover tasks from failures. Currently, most stream processors are transactional processing systems governed by consistency rules and processing guarantees. This section highlights the types of guarantees offered by different stream processing systems and implementation strategies that materialize them.

\para{Past Challenges and The Lambda Architecture: } When large scale computing became  mainstream, a design pattern emerged called ``lambda architecture'' which suggested the separation of systems across different layers according to their specialization and reliability capabilities. Hadoop and transactional databases were reliable in terms of processing guarantees, thus, they could take all critical computation. Whereas, stream processing systems could achieve low latency and scale but they did not offer a clear set of consistency guarantees. For example, in the state-oblivious Storm system the fault-tolerance approach would solely consider which input events have been fully processed or not and which should be replayed on a timeout. Nevertheless, there was no clear picture of what level of consistency can be expected from stream processors. At the same time, databases had formal guarantees. For example, a set of transactions would be processed using ACID guarantees, which includes atomicity across transactions, consistency for the valid states a database can have, isolation in terms of concurrent execution, and durability on what can be recovered after failure. To reason about consistency in the context of data streaming, there had been a need to lay out a set of assumptions (e.g., logged input) and processing granularity for defining a concept related to transactions.

\stitle{Consistent State in Stream Processing.} A stream processor today is a distributed system consisting of different concurrently executing tasks. Source tasks subscribe to input streams that are typically recorded in a partitioned log such as Kafka and therefore input streams can be replayed. Sink tasks commit output streams to the outside world and  every task in this system can contain its own state. For example, source tasks need to keep the current position of their input streams in their state. 
A system execution can be often modeled through the concept of ``concurrent actions''. An action includes: invoking stream task logic on an input event, mutating its state, and producing output events. Every action happening in such a system causes other actions. Effectively, just a single record sent by a source contributes to state updates throughout the whole pipeline and output events created by the sinks. If a specific action is lost or happens twice, then the complete system enters into an \textit{inconsistent} state.

Fault tolerance is an integral aspect of streaming systems that significantly impacts their state consistency. We analyze the fault tolerance strategies of existing streaming systems in Section~\ref{sub:fault-tolerance}.
In addition, due to causal dependencies on state, the order of action execution is also critical. Existing reliable stream processors either define a transaction out of each action or a coarse grained set of actions that we call \emph{epochs}. We explain these approaches in more detail, next.

\subsubsection{State Persistence at Event Granularity}
A form of consistent processing in data streaming is employing a transaction per local action. Google's Millwheel, the cloud runtime for the dataflow data streaming service, employs such a strategy. Millwheel uses BigTable to commit each full compute action which includes: input events,  state transitions and generated output. The act of committing these actions is also called a ``strong production'' in Millwheel. 

Persisting state of an operator per output event, is an approach which seemingly induces high latency overhead. However, traditional database optimizations can be used to speed up commit and state read times. Write ahead logging, blind writes, bloom filters, and batch commits at the storage layer can be used to reduce the commit latency. More importantly, since the order of actions is predefined at commit time, state persistence on a per-event basis also guarantees deterministic executions. In addition, this approach has important effects on consistency as perceived by applications that consume the system's output. This follows from the fact that ``exactly-once processing'' in this context relates to each action being atomically committed, as we discuss in Section~\ref{subsub:output-commit}.

\begin{figure}[t]
        \centering
        \includegraphics[width=.8\linewidth]{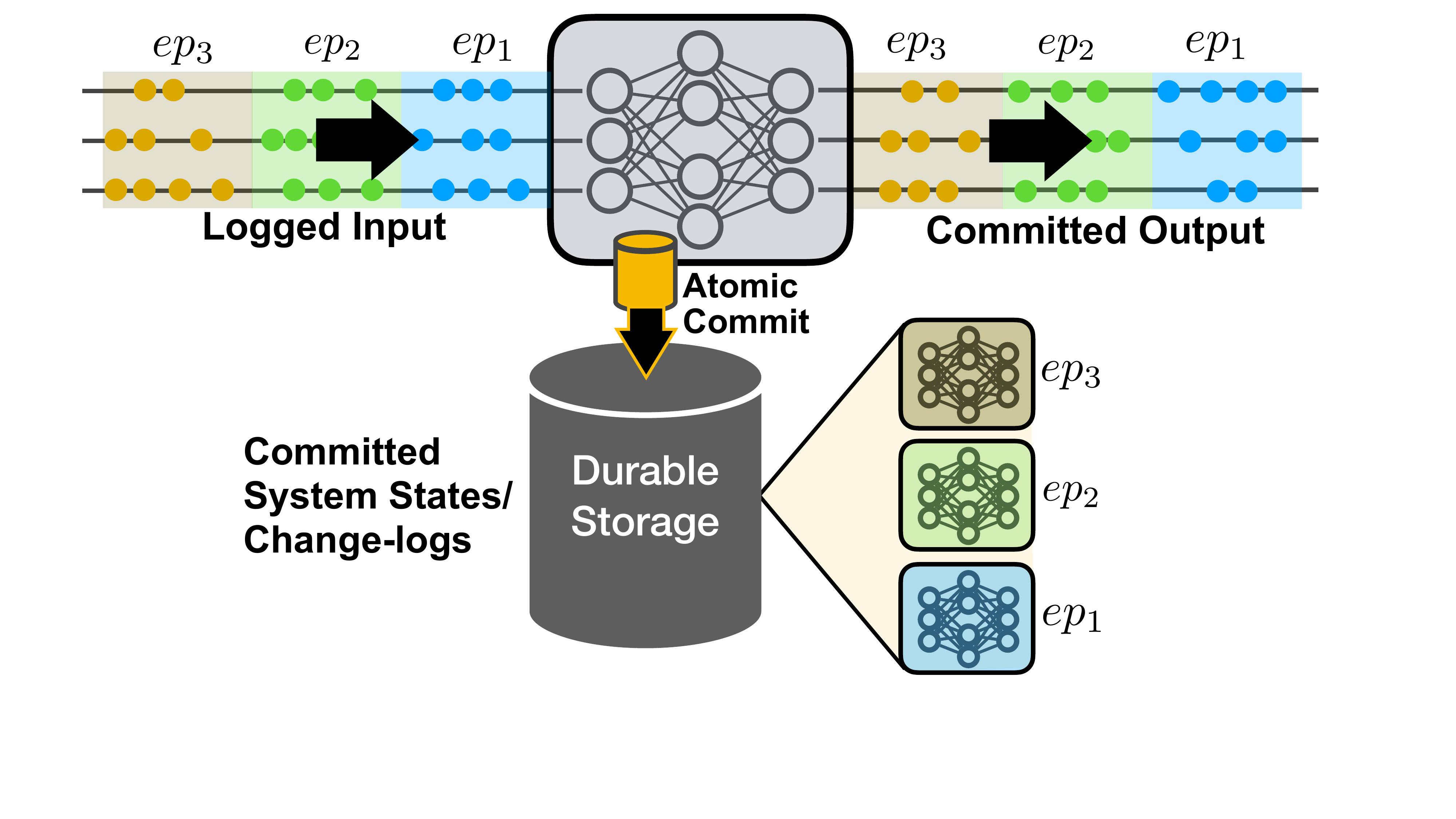}
        \caption{Transactional Epoch Commits in Data Streaming}
        \label{fig:epochcommits}
\end{figure}

\subsubsection{State Persistence at Epoch Granularity}
Instead of adopting state persistence on a per-record granularity processing, epoch-level approaches divide computation into a series of mini-batches, also known as ``epochs''.

In \autoref{fig:epochcommits} we depict the overall approach, marking input, system states and outputs with a distinct epoch identifier. Epochs can be defined through markers at the logged input of the streaming application. A system execution can be instrumented to process each epoch and commit the state of the entire task graph after each epoch is processed. If a failure or other reconfiguration action happens during the execution of an epoch then the system can roll back to a previously committed epoch and recover its execution. The term ``exactly-once processing'' in this context relates to each epoch being atomically committed. In Section~\ref{sub:fault-tolerance} where we present the different  levels of processing semantics in streaming we call this flavor \textit{exactly-once processing on state}. The rest of this section focuses on various approaches used to commit stream epochs.

\stitle{Strict Two-Phase Epoch Commits.} A common coordinated protocol to commit epochs is a strict two-phase commit where: Phase-1 corresponds to the full processing of an epoch and the Phase-2 ensures persisting the state of the system at the end of the computation. 

This approach was popularized by Apache Spark \cite{zaharia2012discretized} through the use of periodic ``micro-batching'' and it is an effective strategy when batch processing systems are used for unbounded processing. The main downside of this approach is the risk of low task utilization due to synchronous execution, since tasks have to wait for all other tasks to finish their current epoch. Drizzle~\cite{venkataramandrizzle} mitigates this problem by chaining multiple epochs in a single atomic commit. A similar approach was also employed by S-Store~\cite{meehan2015s}, where each database transaction corresponds to an epoch of the input stream that is already stored in the same database.

\stitle{Asynchronous Two-Phase Epoch Commits.} 
For pure dataflow systems, strict two-phase committing is problematic since tasks are uncoordinated and long-running. Furthermore, it is feasible to achieve the same functionality asynchronously through consistent snapshotting algorithms, known from classic distributed systems literature \cite{carbone2018scalable}. Consistent snapshotting algorithms exhibit beneficial properties because they not require pausing a streaming application. Furthermore, they acquire a snapshot of a consistent cut in a distributed execution \cite{chandy1985distributed}. In other words, they capture the global states of the system during a ``valid'' execution. Throughout different implementations we can identify i) unaligned and ii) aligned snapshotting protocols.

\para{I. Unaligned / Chandy-Lamport \cite{chandy1985distributed} snapshots} provide one of the most efficient methods to obtain a consistent snapshot. This approach is currently supported by several stream processors, such as IBM Streams and Flink. The core idea is to make use of a punctuation or ``marker'', into the regular stream of events and use that marker to separate all actions that come before and after the snapshot while the system is running. A caveat of unaligned snapshots is the need to record input (a.k.a. in-flight) events that arrive to individual tasks until the protocol is complete. In addition to space overhead for logged inputs, unaligned snapshots require more processing during recovery, since logged inputs need to be replayed (similarly to redo logs in database recovery with fuzzy checkpoints).
 
\para{II. Aligned Snapshots}
Aligned snapshots aim to improve performance during recovery and minimize reconfiguration complexity exhibited by unaligned snapshots. The main differentiation is to prioritize input streams that are expected before the snapshot and thus, end up solely with states that reflect a complete computation of an epoch and no in-flight events as part of a snapshot. For example, Flink's epoch snapshotting mechanism \cite{CarboneEF17,carbone2015lightweight} resembles the Chandy Lamport algorithm in terms of using markers to identify epoch frontiers. However, it additionally employs an alignment phase that synchronizes markers within tasks before disseminating further. This is achieved through partially blocking input channels where markers were previously received until all input channels have transferred all messages corresponding to a particular epoch. 

In summary, unaligned snapshots are meant to offer the best runtime performance but sacrifice recovery times due to the redo-phase needed upon recovery. Whereas, aligned snapshots can lead to slower commit times due to the alignment phase while providing a set of beneficial properties. First, aligned snapshots reflect a complete execution of an epoch which is useful in use-cases where snapshot isolated queries need to be supported on top of data streaming \cite{squery}. Furthermore, aligned snapshots yield the lowest reconfiguration footprint as well as setting the basis for live reconfiguration within the alignment phase as exhibited by Chi \cite{mai2018chi}. 

\subsection{1st vs. 2nd Generation}
State is a concept that has been very central to stream processing. The notion of state itself has been addressed with many names such as ``summary'', ``synopsis'', ``sketch'' or ``stream table'' and it reflects the evolution of data stream management along the years. Early DSMS systems~\cite{babcock2002models,abadi2003aurora,arasu2004stream,chandrasekaran2003telegraphcq} (circa 2000-2010) hinted state and its management from the user. They declared and managed internally in in-memory all data structures needed to support a selected set of operations. This type of state, often referred to as ``summary'' was used to internally materialize continuous processing operators such as those of the time-varying relational model of CQL~\cite{cql}, as seen in STREAM~\cite{arasu2004stream}.

A decade later, scalable data computing systems based on the MapReduce~\cite{dean2004mapreduce} architecture allowed for arbitrary user-defined logic to be scaled and executed reliably using distributed middleware and partitioned file systems. Following the same trend, many existing data management models were revisited and re-architectured with scalability in mind (e.g., NoSQL, NewSQL databases). Similarly, a growing number of scalable data stream processing systems~\cite{carbone2015apache,akidau2015dataflow,akidau2013millwheel,seep} married principles of scalable computing with stream semantics and models that were identified in the past (e.g. out-of-order processing~\cite{srivastava2004flexible,li2008out}). This pivoting helped stream management technology to lift all assumptions associated with limited state capacity and thus reach its nearly full potential of executing correctly continuous event-driven applications with arbitrary state. 

As of today, modern stream processors can compile and execute graphs of long-running operators with complete, user-defined state yet system-managed that is fault-tolerant and reconfigurable given a clear set of transactional guarantees\cite{castro2013integrating,CarboneEF17,akidau2013millwheel}.  

\subsection{Open Problems}

Data streaming covers many data management needs today that go beyond real-time analytics, which was the original purpose of the stream processing technology. New needs include support for more complex data pipelines with implicit transactional guarantees. Furthermore, modern applications involve Machine Learning, Graph Analysis and Cloud Apps, all of which have a common denominator: complex state and new access patterns. These needs have cultivated novel research directions in the emerging field of stream state management.

The decoupling of state programming from state persistence resembles the concept of data independence in databases. Systems are converging in terms of semantics and operations on state while, at the same time many new methods employed on embedded databases (e.g., LSM-trees, state indexing, externalized state) are helping stream processors to evolve in terms of performance capabilities. A recent study~\cite{kalavri2020support} showcases the potential of workload-aware state management, adapting state persistence and access to the individual operators of a dataflow graph. To this end, an increasing number of ``pluggable'' systems~\cite{chandramouli2018faster,zhu2019livegraph} for local state management with varying capabilities are being adopted by stream processors. This opens new capabilities for optimization and sophisticated, yet transparent state management that can automate the process of selecting the right physical plan and reconfigure that plan while continuous applications are executed.

\section{Fault Tolerance \& High Availability}
\label{sec:FT}

\setlength{\tabcolsep}{6pt}
\begin{table*}
\smaller\centering
\caption{Fault-tolerance in streaming systems.}
\label{tab:ha-sps}
{
\begin{tabular}{p{0.13\textwidth}
                        p{0.04\textwidth}
                        p{0.04\textwidth}
                        p{0.03\textwidth}
                        p{0.04\textwidth}
                        p{0.05\textwidth}
                        p{0.03\textwidth}
                        p{0.03\textwidth}
                        p{0.05\textwidth}
                        p{0.03\textwidth}
                        p{0.04\textwidth}
                        p{0.04\textwidth}
                        p{0.09\textwidth}
                        p{0.03\textwidth}
                        }
\hline
\textbf{System} &
\multicolumn{3}{c}{\textbf{Processing semantics}} &
\multicolumn{3}{c}{\textbf{Replication}} &
\multicolumn{3}{c}{\textbf{Recovery data}} & 
\multicolumn{4}{c}{\textbf{Storage medium}} \\
\rowcolor{white}
& Least
& \multicolumn{2}{c}{Exactly-once}
& Active
& Passive
& None
& State
& Output
& None
& \multicolumn{2}{c}{Resilient store}
& In-Memory
& None \\

&
& State
& Output
&
&
&
&
&
&
& Local
& Remote
&
&
\\
\hline

Aurora*~\cite{cherniack2003scalable}
& \multicolumn{1}{a}{\checkmark} 
& \multicolumn{1}{a}{} 
& \multicolumn{1}{a}{}
& \multicolumn{1}{b}{} 
& \multicolumn{1}{b}{\checkmark}
& \multicolumn{1}{b}{} 
& \multicolumn{1}{a}{} 
& \multicolumn{1}{a}{\checkmark}
& \multicolumn{1}{a}{} 
& \multicolumn{1}{b}{} 
& \multicolumn{1}{b}{} 
& \multicolumn{1}{b}{\checkmark}
& \multicolumn{1}{b}{} 
\\
\hline
TelegraphCQ~\cite{ShahHB04}
& \multicolumn{1}{a}{} 
& \multicolumn{1}{a}{\checkmark} 
& \multicolumn{1}{a}{}
& \multicolumn{1}{b}{\checkmark}
& \multicolumn{1}{b}{} 
& \multicolumn{1}{b}{} 
& \multicolumn{1}{a}{\checkmark}
& \multicolumn{1}{a}{\checkmark}
& \multicolumn{1}{a}{} 
& \multicolumn{1}{b}{\checkmark}
& \multicolumn{1}{b}{} 
& \multicolumn{1}{b}{} 
& \multicolumn{1}{b}{} 
\\
\hline

Borealis~\cite{abadi2005design, balazinska2008fault}
& \multicolumn{1}{a}{\checkmark}
& \multicolumn{1}{a}{} 
& \multicolumn{1}{a}{} 
& \multicolumn{1}{b}{\checkmark}
& \multicolumn{1}{b}{} 
& \multicolumn{1}{b}{} 
& \multicolumn{1}{a}{\checkmark}
& \multicolumn{1}{a}{\checkmark}
& \multicolumn{1}{a}{} 
& \multicolumn{1}{b}{} 
& \multicolumn{1}{b}{} 
& \multicolumn{1}{b}{\checkmark}
& \multicolumn{1}{b}{} 
\\
\hline

%
S4~\cite{neumeyer2010s4}
& \multicolumn{1}{a}{\checkmark}
& \multicolumn{1}{a}{} 
& \multicolumn{1}{a}{} 
& \multicolumn{1}{b}{} 
& \multicolumn{1}{b}{} 
& \multicolumn{1}{b}{\checkmark}
& \multicolumn{1}{a}{} 
& \multicolumn{1}{a}{} 
& \multicolumn{1}{a}{\checkmark}
& \multicolumn{1}{b}{} 
& \multicolumn{1}{b}{} 
& \multicolumn{1}{b}{} 
& \multicolumn{1}{b}{\checkmark}
\\
\hline

Seep~\cite{castro2013integrating,fernandez2014making}
& \multicolumn{1}{a}{} 
& \multicolumn{1}{a}{} 
& \multicolumn{1}{a}{\checkmark}
& \multicolumn{1}{b}{} 
& \multicolumn{1}{b}{\checkmark}
& \multicolumn{1}{b}{} 
& \multicolumn{1}{a}{\checkmark}
& \multicolumn{1}{a}{\checkmark}
& \multicolumn{1}{a}{} 
& \multicolumn{1}{b}{} 
& \multicolumn{1}{b}{\checkmark}
& \multicolumn{1}{b}{} 
& \multicolumn{1}{b}{} 
\\
\hline

%
Naiad~\cite{murray2013naiad}
& \multicolumn{1}{a}{} 
& \multicolumn{1}{a}{\checkmark}
& \multicolumn{1}{a}{} 
& \multicolumn{1}{b}{} 
& \multicolumn{1}{b}{\checkmark}
& \multicolumn{1}{b}{} 
& \multicolumn{1}{a}{\checkmark}
& \multicolumn{1}{a}{\checkmark}
& \multicolumn{1}{a}{} 
& \multicolumn{1}{b}{} 
& \multicolumn{1}{b}{\checkmark}
& \multicolumn{1}{b}{} 
& \multicolumn{1}{b}{} 
\\
\hline

Timestream~\cite{QianHS13}
& \multicolumn{1}{a}{} 
& \multicolumn{1}{a}{} 
& \multicolumn{1}{a}{\checkmark}
& \multicolumn{1}{b}{} 
& \multicolumn{1}{b}{\checkmark}
& \multicolumn{1}{b}{} 
& \multicolumn{1}{a}{\checkmark}
& \multicolumn{1}{a}{\checkmark}
& \multicolumn{1}{a}{} 
& \multicolumn{1}{b}{} 
& \multicolumn{1}{b}{\checkmark}
& \multicolumn{1}{b}{} 
& \multicolumn{1}{b}{} 
\\
\hline

Millwheel~\cite{akidau2013millwheel}
& \multicolumn{1}{a}{} 
& \multicolumn{1}{a}{} 
& \multicolumn{1}{a}{\checkmark}
& \multicolumn{1}{b}{} 
& \multicolumn{1}{b}{\checkmark}
& \multicolumn{1}{b}{} 
& \multicolumn{1}{a}{\checkmark}
& \multicolumn{1}{a}{\checkmark}
& \multicolumn{1}{a}{} 
& \multicolumn{1}{b}{} 
& \multicolumn{1}{b}{\checkmark}
& \multicolumn{1}{b}{} 
& \multicolumn{1}{b}{} 
\\
\hline

%
Storm~\cite{toshniwal2014storm}
& \multicolumn{1}{a}{\checkmark}
& \multicolumn{1}{a}{} 
& \multicolumn{1}{a}{} 
& \multicolumn{1}{b}{} 
& \multicolumn{1}{b}{} 
& \multicolumn{1}{b}{\checkmark}
& \multicolumn{1}{a}{} 
& \multicolumn{1}{a}{} 
& \multicolumn{1}{a}{\checkmark}
& \multicolumn{1}{b}{} 
& \multicolumn{1}{b}{} 
& \multicolumn{1}{b}{} 
& \multicolumn{1}{b}{\checkmark}
\\
\hline

Trident~\cite{CUSTOM:web/trident}
& \multicolumn{1}{a}{} 
& \multicolumn{1}{a}{} 
& \multicolumn{1}{a}{\checkmark}
& \multicolumn{1}{b}{} 
& \multicolumn{1}{b}{\checkmark}
& \multicolumn{1}{b}{} 
& \multicolumn{1}{a}{\checkmark}
& \multicolumn{1}{a}{} 
& \multicolumn{1}{a}{} 
& \multicolumn{1}{b}{} 
& \multicolumn{1}{b}{\checkmark}
& \multicolumn{1}{b}{} 
& \multicolumn{1}{b}{} 
\\
\hline

%
S-Store~\cite{CetintemelDK14, TatbulZM15}
& \multicolumn{1}{a}{} 
& \multicolumn{1}{a}{\checkmark}
& \multicolumn{1}{a}{} 
& \multicolumn{1}{b}{} 
& \multicolumn{1}{b}{\checkmark}
& \multicolumn{1}{b}{} 
& \multicolumn{1}{a}{\checkmark}
& \multicolumn{1}{a}{} 
& \multicolumn{1}{a}{} 
& \multicolumn{1}{b}{\checkmark}
& \multicolumn{1}{b}{} 
& \multicolumn{1}{b}{} 
& \multicolumn{1}{b}{} 
\\
\hline

Trill~\cite{chandramouli2014trill}
& \multicolumn{1}{a}{} 
& \multicolumn{1}{a}{\checkmark}
& \multicolumn{1}{a}{} 
& \multicolumn{1}{b}{} 
& \multicolumn{1}{b}{\checkmark}
& \multicolumn{1}{b}{} 
& \multicolumn{1}{a}{\checkmark}
& \multicolumn{1}{a}{} 
& \multicolumn{1}{a}{} 
& \multicolumn{1}{b}{\checkmark}
& \multicolumn{1}{b}{} 
& \multicolumn{1}{b}{} 
& \multicolumn{1}{b}{} 
\\
\hline

Heron~\cite{KulkarniBF15}
& \multicolumn{1}{a}{\checkmark}
& \multicolumn{1}{a}{} 
& \multicolumn{1}{a}{} 
& \multicolumn{1}{b}{} 
& \multicolumn{1}{b}{} 
& \multicolumn{1}{b}{\checkmark}
& \multicolumn{1}{a}{} 
& \multicolumn{1}{a}{} 
& \multicolumn{1}{a}{\checkmark}
& \multicolumn{1}{b}{} 
& \multicolumn{1}{b}{} 
& \multicolumn{1}{b}{} 
& \multicolumn{1}{b}{\checkmark}
\\
\hline

Streamscope~\cite{LinHZ16}
& \multicolumn{1}{a}{} 
& \multicolumn{1}{a}{} 
& \multicolumn{1}{a}{\checkmark} 
& \multicolumn{1}{b}{\checkmark}
& \multicolumn{1}{b}{\checkmark}
& \multicolumn{1}{b}{\checkmark}
& \multicolumn{1}{a}{\checkmark} 
& \multicolumn{1}{a}{} 
& \multicolumn{1}{a}{\checkmark} 
& \multicolumn{1}{b}{} 
& \multicolumn{1}{b}{\checkmark}
& \multicolumn{1}{b}{} 
& \multicolumn{1}{b}{\checkmark}
\\
\hline

Streams~\cite{jacques2016consistent}
& \multicolumn{1}{a}{} 
& \multicolumn{1}{a}{\checkmark} 
& \multicolumn{1}{a}{} 
& \multicolumn{1}{b}{} 
& \multicolumn{1}{b}{\checkmark}
& \multicolumn{1}{b}{} 
& \multicolumn{1}{a}{\checkmark}
& \multicolumn{1}{a}{} 
& \multicolumn{1}{a}{} 
& \multicolumn{1}{b}{} 
& \multicolumn{1}{b}{\checkmark}
& \multicolumn{1}{b}{} 
& \multicolumn{1}{b}{} 
\\
\hline

Samza~\cite{NoghabiPP17}
& \multicolumn{1}{a}{\checkmark}
& \multicolumn{1}{a}{} 
& \multicolumn{1}{a}{} 
& \multicolumn{1}{b}{} 
& \multicolumn{1}{b}{\checkmark}
& \multicolumn{1}{b}{} 
& \multicolumn{1}{a}{\checkmark} 
& \multicolumn{1}{a}{} 
& \multicolumn{1}{a}{} 
& \multicolumn{1}{b}{\checkmark}
& \multicolumn{1}{b}{} 
& \multicolumn{1}{b}{} 
& \multicolumn{1}{b}{} 
\\
\hline

Flink~\cite{CarboneEF17, carbone2015apache}
& \multicolumn{1}{a}{} 
& \multicolumn{1}{a}{\checkmark}
& \multicolumn{1}{a}{} 
& \multicolumn{1}{b}{} 
& \multicolumn{1}{b}{\checkmark}
& \multicolumn{1}{b}{} 
& \multicolumn{1}{a}{\checkmark}
& \multicolumn{1}{a}{} 
& \multicolumn{1}{a}{} 
& \multicolumn{1}{b}{} 
& \multicolumn{1}{b}{\checkmark}
& \multicolumn{1}{b}{} 
& \multicolumn{1}{b}{} 
\\
\hline

Spark~\cite{armbrust2018structured}
& \multicolumn{1}{a}{} 
& \multicolumn{1}{a}{\checkmark} 
& \multicolumn{1}{a}{} 
& \multicolumn{1}{b}{} 
& \multicolumn{1}{b}{\checkmark}
& \multicolumn{1}{b}{} 
& \multicolumn{1}{a}{\checkmark}
& \multicolumn{1}{a}{} 
& \multicolumn{1}{a}{} 
& \multicolumn{1}{b}{} 
& \multicolumn{1}{b}{\checkmark}
& \multicolumn{1}{b}{} 
& \multicolumn{1}{b}{} 
\\
\hline

\end{tabular}
}
\end{table*}

Fault tolerance is a system's capacity to continue its operation in spite of failures delivering the expected service as if no failures had happened.
It is specially important for streaming systems for two reasons.
First, streaming systems conduct stateful computations over potentially unbounded data streams.
Without fault tolerance streaming systems would have to redo computations from the beginning given that the state or progress thus far would be lost during a failure.
Besides losing processing progress accumulated over an arbitrary time period, recomputation is many times infeasible because the already processed segment of a data stream has permanently vanished.

Second, contemporary streaming systems feature a distributed systems architecture for scalability.
In a system deployed on multiple physical machines failures occur commonly.
Based on this motivation, a lot of exciting work has been performed on fault tolerance in streaming systems.
We present it in Section~\ref{sub:fault-tolerance}.

In computer systems, availability is defined as the time period that a system accomplishes its service relative to service interruption periods.
It is typically quantified as a percentage, 100\% being perfect availability~\cite{Gray91High}.
The term high availability has been adopted to denote that a system achieves a very high percentage of availability like 99.999\% or higher.

In stream processing where systems are not probed by users as in the case of typical information systems like web applications, what service accomplishment means is open to interpretation.
Surprisingly, no definition for high availability is provided in the stream processing literature.
Existing research (Section~\ref{sub:high-availability}) quantifies high availability using combinations of three metrics, namely recovery time, performance overhead in terms of throughput and latency, and resource utilization.
We highlight the absence of a definition and suitable metric for high availability in the open problems in Section~\ref{sub:ha-open-problems} where we propose a definition based on processing progress and a proxy for measuring high availability based on end-to-end latency.
Before finishing with the open problems, we separate the 1st generation from the modern in fault tolerance and high availability in Section~\ref{sub:ft-ha-vintage}.

\subsection{Fault-tolerance}
\label{sub:fault-tolerance}

Many important challenges in stream processing manifest when we take into account failures.
Managing failures in a distributed streaming system entails maintaining snapshots of state, migrating state, and scaling out operators while affecting as least as possible the healthy parts of the system.
Table~\ref{tab:ha-sps} presents the fault-tolerance strategies of
eighteen streaming systems arranged in order of publication appearance
from past to present.
We analyse the strategies across the following four dimensions.

\textit{1. Processing semantics} conveys how a system's data processing is affected by failures.
Typically, all systems in the literature are able to produce correct results in failure-free executions.
But to mask a failure completely is hard especially in the stream processing domain where, typically, output is delivered as soon as it is produced. 

In recent years the stream processing domain has settled on the terms \textit{at least-once} and \textit{exactly-once} to characterize the processing semantics~\cite{armbrust2018structured, carbone2015apache, jacques2016consistent, LinHZ16, NoghabiPP17}.
At most-once is also part of the nomenclature but it is mostly obsolete as systems opt to support one of the two stronger levels.
At least-once processing semantics means that the system will produce the same results as a failure-free execution with the addition of duplicate records as a side effect of recovery.

Exactly-once lends itself to two different interpretations.
A system may support exactly-once processing semantics within its boundaries ensuring that any inconsistencies or duplicate execution carried out on recovery is not part of its state.
We call that exactly-once processing semantics on \textit{state}.
It should be noted that most systems in this category still assume that the computations they apply as well as the system's functions are deterministic, which is often not the case; processing-time windows and operators processing input from multiple sources are two prime examples of nondeterminism. With nondeterminism at play, the system's state on recovery can diverge. Clonos~\cite{silvestre2021clonos} provides exactly-once processing including nondeterministic computations by means of causal consistency. It keeps determinants about nondeterministic computations in a resilient manner and uses them to regenerate the exact computational state following a failure.

While a system can restore its state to a consistent snapshot, the same is not feasible in general to accomplish with the output published by the system.
Once the output is out, it is available for consumption by external applications.
Thus, a system with exactly-once processing semantics on state will still produce duplicate output on recovery.
This problem has been termed the output commit problem~\cite{ElnozahyAW02} in the distributed systems literature.
Systems that manage to produce the same output under failure as a failure-free execution have exactly-once processing semantics on \textit{output}.
In Section~\ref{subsub:output-commit} we elaborate how streaming systems treat the output commit problem.

\noindent \textit{2. Replication} regards the use of additional computational resources for recovering an execution.
We adopt the terminology of Hwang et al.~\cite{HwangBR05} that classify replication as either \textit{active} where two instances of the same execution run in parallel or \textit{passive} where each running stateful operator that is part of an execution dispatches its checkpointed state to a standby operator.

\noindent \textit{3. Recovery data} addresses what data are regularly stored for recovery purposes.
Data may include the \textit{state} of each operator and the \textit{output} it produces.
In addition, many fault tolerance strategies need to replay tuples of input streams during recovery in order to reprocess them.
For this purpose input streams are persistently stored typically in message brokers like Apache Kafka.
However, we exclude this fact from the table to save space.

\noindent \textit{4. Storage medium} states where recovery data is stored.
It can be in a \textit{resilient store} that is \textit{local} to each stateful operator, in a \textit{remote} resilient store, or in the memory space of a stateful operator.
\textit{In-memory} means that operators use their memory space as a primary storage medium for recovery data.
Systems that cache data for recovery in memory like output tuples do not fall in this category.

The table is meant to be read both horizontally to describe a specific system's approach to fault tolerance and vertically to uncover how the different building blocks shape the landscape of fault tolerance in stream processing.
Two remarks are necessary.
First, the table contains three more annotations besides the self-explanatory checkmarks.
Streamscope~\cite{LinHZ16} presents and evaluates three distinct fault tolerance strategies, an active replication-based strategy , a passive one, and a strategy that relies on recomputing state by replaying data from input streams.
Second, the state column in the recovery data dimension captures not only checkpointed state but also state metadata that allow recomputing the state, such as a changelog~\cite{NoghabiPP17} or state dependencies~\cite{QianHS13}.

The table reveals four interesting patterns.
First, of all columns, two accumulate the majority of checkmarks, passive replication and storing state for recovery.
This is perhaps the most visible pattern on the table that signifies that passive replication by storing state is, unsurprisingly, a very popular option for streaming systems.
One typical recovery approach is to restore the latest checkpoint of a failed operator in a new node and replay input that appeared after the checkpoint.
Variations of this approach include saving in-flight tuples along with the state and maintaining in-flight tuples in upstream nodes.
Second, storing in-flight tuples for recovery is not preferred anymore, although it was a popular option for streaming systems in the past.
Third, while past systems strived to support exactly-once output processing semantics, later systems opt for exactly-once semantics on state and outsource the deduplication of output to external systems.
We will elaborate on this aspect in Section~\ref{subsub:output-commit}. 
Finally, among the various storage media for recovery data a remote resilient store is the clear winner.

\subsubsection{The output commit problem}
\label{subsub:output-commit}


The output commit problem~\cite{ElnozahyAW02} 
specifies that a system should only publish output to the outside world when it is certain that it can recover the state from where the output was published so that every output is only published once because output cannot be retracted once it is sent.
If output is sent twice, then the system manifests inconsistent behavior with respect to the outside world.
An important instance of this problem manifests when a system is restoring some previous consistent state due to a failure.
In contrast to the system's state, its output cannot be retracted in general.
Thus, under failures, systems must be careful not to produce duplicate output.

The output commit problem is relevant in streaming systems, which typically conform to a distributed architecture and process unbounded data streams.
In this setting, the side effects of failures are difficult to mask. 
Streaming systems that solve the output commit problem provide output exactly-once.
Other terms that refer to the same problem are processing output exactly-once and
its paraphrases, as well as precise recovery~\cite{HwangBR05} and strong productions~\cite{akidau2013millwheel}.

Although the problem is relevant and hard, solutions in the stream processing domain are scattered in the literature pertaining to each system in isolation.
We group the various solutions in three categories, transaction-based, progress-based, and lineage-based, and describe each noting the assumptions it involves.
Each of the three types of techniques, use a different trait of the input or computation, to identify
whether a certain tuple has appeared again. Transaction-based techniques use tuple identity, progress-based techniques
use order, while lineage-based techniques use input-output dependencies.
Finally, we provide two more categories of solutions, special sink operators and external sinks that do solve the problem practically, but strictly speaking they do not meet the problem’s specification because they are either specific or external to a streaming system.

\para{Transaction-based.}
Millwheel~\cite{akidau2013millwheel} and Trident~\cite{CUSTOM:web/trident} rely on committing unique ids with records to eliminate duplicate retries.
Millwheel assigns a unique id to each record entering the system and commits every record it produces to a highly available storage system before sending it downstream.
Downstream operators acknowledge received records.
If a delivered record is retried it is ignored by checking the unique id that it carries.
Millwheel assumes no input ordering or determinism.
Trident, on the other hand, batches records into a transaction, which is assigned a unique transaction id and applies a state update to the state backend.
Assuming that transactions are ordered, Trident can accurately ignore retried batches by checking the transaction id.

\para{Progress-based.}
Seep~\cite{fernandez2014making} uses timestamp comparison to deliver output exactly-once relying on the order of timestamps.
Each operator generates increasing scalar timestamps and attaches them to records.
Seep checkpoints the state and output of each operator together with the vector timestamps of the latest records from each upstream operator that affected the operator's state.
On recovery, the latest checkpoint is loaded to a new operator, which replays the checkpointed output records and processes replayed records sent by its upstream operators.
Downstream operators discard duplicate records based on the timestamps.
The system assumes deterministic computations that do not rely on system time or random input.

A previous version of Seep~\cite{castro2013integrating} applies the same process with the difference that a recovered operator rewinds its logical clock to the timestamp of the checkpoint it possesses before emitting records.
The system assumes deterministic computations without side-effects and a monotonically increasing logical clock providing timestamps.
It further assumes that records in a stream are ordered by their timestamps.

\para{Lineage-based.}
Timestream~\cite{QianHS13} and Streamscope~\cite{jacques2016consistent} use dependency tracking to provide exactly-once output.
During normal operation, both systems track operator input and output dependencies by uniquely identifying records with sequence numbers.
Streamscope persists records with their identifiers asynchronously.
Both systems store operator dependencies periodically in an asynchronous manner.
In Streamscope, however, each operator checkpoints individually not only its dependencies but also its state.
On recovery, Timestream retrieves the dependencies of failed operators by contacting upstream nodes recursively until all inputs required to rebuild the state are made available.
Streamscope follows a similar process, but starts from a failed operator's checkpoint snapshot.
For each input sequence number in that snapshot not found in persistent storage Streamscope contacts upstream operators, which may have to recompute the record starting from their most relevant snapshot that can produce the output record given its sequence number.
Finally, both systems use garbage collection to discard obsolete dependencies but in a subtly different manner.
Timestream computes the input records required by upstream operators in reverse topological order from the final output to the original input and discards those unneeded.
Streamscope does the same but instead of computing dependencies, it uses low watermarks per operator and per stream to discard snapshots and records that are behind.
In Timestream storing dependencies asynchronously can lead to duplicate recomputation, but downstream operators bearing the correct set of dependencies can discard them.
Streamscope applies the same process only if duplicate records cannot be found in persistent storage.
Both Timestream and Streamscope assume deterministic computation and input in terms of order and values.

The time-based and lineage-based solutions are vulnerable to failures of the last operator(s) on the dataflow graph, which produce the final output, since both solutions rely on downstream operators for filtering duplicate records.

\begin{table}[t]
  \caption{Assumptions that systems make for solving the output commit problem}
  \label{tab:output-commit}
  \begin{tabular}{p{0.09\textwidth}p{0.35\textwidth}}
      \textbf{System} & \textbf{Assumptions} \\
      \hline
      Millwheel & \multicolumn{1}{e}{External High-throughput Transactional Store} \\ \hline
      Timestream & \multicolumn{1}{e}{Deterministic computation and input} \\ \hline
      Streamscope & \multicolumn{1}{e}{Deterministic computation and input} \\ \hline
      Trident & \multicolumn{1}{e}{Deterministic computation and input, ordering of} \\
      & \multicolumn{1}{e}{transactions} \\ \hline
      Seep & \multicolumn{1}{e}{Deterministic computation, monotonically-} \\
      & \multicolumn{1}{e}{increasing logical clock, records ordered by} \\
      & \multicolumn{1}{e}{timestamp} \\
      \hline
  \end{tabular}
\end{table}

\para{Special sink operators.}
Streams~\cite{jacques2016consistent} implements special sinks for retracting output from files and databases.
The application of this approach solves the output commit problem for specific use cases, but it is not applicable in general since it defies the core assumption of the problem that output cannot be retracted.

\para{External sinks.}
Some systems like Streams~\cite{jacques2016consistent}, Flink~\cite{carbone2015apache}, and Spark~\cite{armbrust2018structured} provide exactly-once semantics on state and outsource the output commit problem to external sinks that support idempotent writes, such as Apache Kafka.

One way to categorise the solutions provided by special sink operators and external sinks, is as optimistic output
techniques, that push output immediately and retract it or update it if needed, and pessimistic output techniques that use
a form of write ahead log, to write the output they will publish, if everything goes well until the output is permanently committed~\cite{CarboneEF17}.
Optimistic output techniques, which resemble multi-version concurrency control from the database world, include modifiable
and versioned output destinations, while pessimistic output techniques include transactional sinks and similar tools.


\subsection{High availability}
\label{sub:high-availability}


Empirical studies of high availability in stream processing~\cite{HwangBR05} propose an active replication approach~\cite{balazinska2008fault, ShahHB04}, a passive replication approach~\cite{GuZY09, HwangX07, KwonB08}, a hybrid active-passive replication approach~\cite{HeinzeZ15, SuZ16, ZhangGY10}, or model multiple approaches and evaluate them with simulated experiments~\cite{ChandramouliG17, HwangBR05}.


\para{Active replication.}
Flux~\cite{ShahHB04} implements active replication by duplicating the computation and coordinating the progress of the two replicas.
Flux restores operator state and in-flight data of a failed partition while the other partition continues to process input.
A new primary dataflow that runs following a failure quiesces when a new secondary dataflow is ready in a standby machine in order to copy the state of its operators to the new secondary.
Contrastingly, Borealis~\cite{balazinska2008fault} has nodes address upstream node failures by switching to a live replica of the failed upstream node.
If a replica is not available, the node can produce tentative output for incomplete input to avoid the recovery delay.
The approach sacrifices consistency to optimize availability, but guarantees eventual consistency.

\para{Passive replication.}
Hwang et al.~\cite{HwangX07} propose that a server in a cluster has another server as backup where it ships independent parts of its checkpointed state.
When a node fails, its backup servers that hold parts of its checkpointed state initiate recovery in parallel by starting to execute the operators of the failed node whose state they have and collecting the input tuples they have missed from the checkpointed state they possess.
SGuard~\cite{KwonB08} and Clonos~\cite{silvestre2021clonos} save computational resources in another way by checkpointing state asynchronously to a distributed file system.
Upon a failure a node is selected to run a failed operator.
The operator's state is loaded from the file system and its in-memory state is reconstructed before it can join the job.
Beyond asynchronous checkpointing,
a new checkpoint mechanism~\cite{GuZY09} preserves output tuples until an acknowledgment is received from all downstream operators.
Next, an operator trims its output tuples and takes a checkpoint.
The authors show that passive replication still requires longer recovery time than active replication, but with 90\% less overhead due to reduced checkpoint size.

\para{Hybrid replication.}
Zwang et al.~\cite{ZhangGY10} propose a hybrid approach to replication, which operates in passive mode under normal operation, but switches to active mode using a suspended pre-deployed secondary copy when a transient failure occurs.
According to the provided experiment results, their approach saves 66\% recovery time
compared to passive replication
and produces 80\% less message overhead than active replication.
Alternatively, Heinze et al.~\cite{HeinzeZ15} propose to dynamically choose the replication scheme for each operator, either active replication or upstream backup, in order to reduce the recovery overhead of the system by limiting the peak latency under failure below a threshold.
Similarly, Su et al.~\cite{SuZ16} counter correlated failures by passively replicating processing tasks except for a dynamically selected set that is actively replicated.

\para{Modeling and simulations.}
In their seminal work Hwang et al.~\cite{HwangBR05}
model and evaluate the recovery time and runtime overhead of four recovery approaches, active standby, passive standby, upstream backup, and amnesia, across different types of query operators.
The simulated experiments suggest that active standby achieves near-zero recovery time at the expense of high overhead in terms of resource utilization, while passive standby produces worse results in terms of both metrics compared to active standby.
However, passive standby poses the only option for arbitrary query networks.
Upstream backup has the lowest runtime overhead at the expense of longer recovery time.
With a similar goal, Shrink~\cite{ChandramouliG17}, a distributed systems emulator,
evaluates the models of five different resiliency strategies
with respect to uptime \textsc{sla} and resource reservation.
The strategies differ across three axes, single-node vs multi-node, active vs passive replication, and checkpoint vs replay.
According to the experiments with real queries on real advertising data using
Trill~\cite{chandramouli2014trill}, active replication with periodic checkpoints
is proved advantageous in many streaming workloads,
although no single strategy is appropriate for all of them.

\subsection{1st generation vs. 2nd generation}
\label{sub:ft-ha-vintage}

In the early years streaming systems put emphasis on high availability setups with preference towards active replication.
Contrastingly modern systems tend to leverage passive replication especially by allocating extra resources on demand that is appropriate for Cloud setups.
In addition, past systems provided approximate results, while modern systems maintain exactly-once processing semantics over their state under failures.
Although past systems lacked in terms of consistency, mainly due to state management aspects, they strived to solve the output commit problem.
Instead, a typical avenue for modern systems that gains traction is to outsource the deduplication of output to external systems.
Finally, while streaming systems used to store their output in order to be able to replay tuples to downstream operators recovering from a failure, now systems rely increasingly on replayable input source for replaying input subsets.

\subsection{Open Problems}
\label{sub:ha-open-problems}

Many problems wait to be solved in the scope of fault tolerance and high availability in streaming systems.
Three of them include novel solutions to the output commit problem, defining and measuring availability in stream processing, and configuring availability for different application requirements.

First, the importance of the output commit problem has the prospect to increase as streaming systems are used in novel ways like for running event-driven applications. Although we presented five different types of solutions, these suffer from computational cost, strong assumptions, limited applicability, and freshness of output results.
New types of solutions are required that score better in these dimensions.

Second, the literature of high availability in stream processing has significantly enhanced the availability of streaming systems throughout the years.
But, to the best of our knowledge, there has been scant research on what availability means in the area of stream processing.
The generic definition of availability for computer  systems by Gray et al.~\cite{Gray91High} relates availability merely to failures.
According to the definition a system is available when it responds to requests with correct results, which is termed as service accomplishment.
In streaming however, processing is continuous and potentially unbounded.
Responding with correct results becomes more challenging.

The factors that may impair availability in streaming include software and hardware failures, overload, backpressure, and types of processing stall, like checkpoints, state migration, garbage collection, and calls to external systems.
The common denominator of those factors, is that the system falls behind input.
This may not be a problem for other types of systems, like databases which can respond to queries with the historical data
they keep, but streaming systems have to continuously catch up processing with the input in order to provide correct results, that is, in order to be available.

\begin{figure}[t]
  \centering
      \includegraphics[scale=.3]{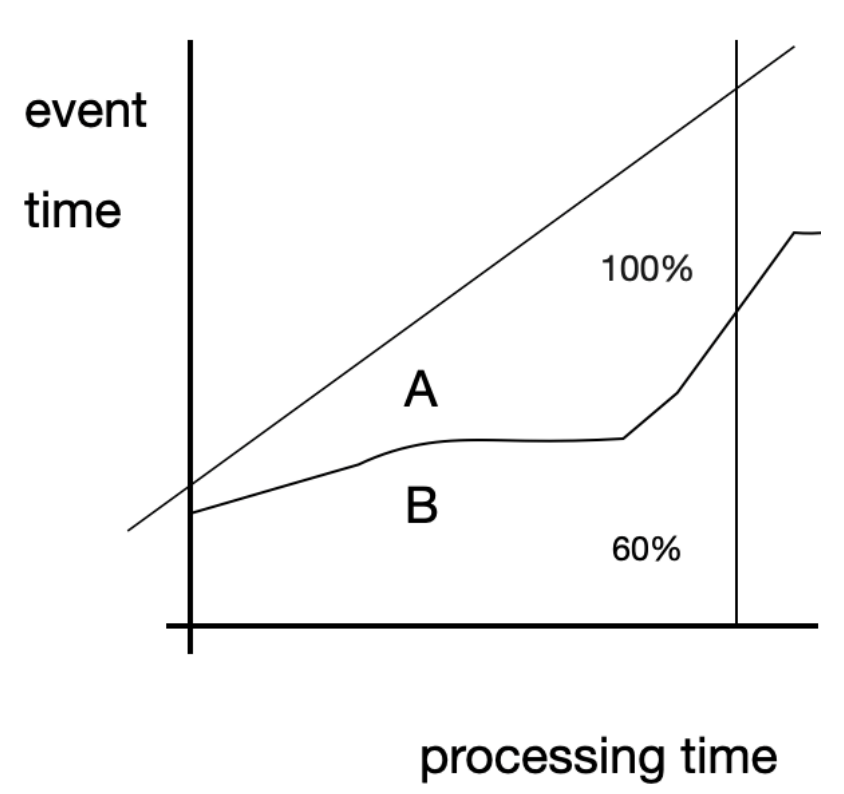}
      \caption{Measuring availability with the slack between processing time and event time over time}
      \label{fig:availability}
\end{figure}

Thus, a more specific definition of availability for stream processing can be  stated in the following way.
\textit{A streaming system is available when it  can  provide output based on the processing of its current input.}
This definition extends to how we measure availability.
An appropriate way would be via progress tracking mechanisms, such as \textit{the slack between processing time and
event time over time}, which quantifies the system’s processing progress with respect to the input as per Figure~\ref{fig:availability}.
The area in the plot signifies the slack between event time and processing time over time.
The surface enclosing A amounts to 100\% availability, while the surface containing B equals 60\% availability.

Last, availability is a prime non-functional characteristic of a streaming system and non-trivial to reason about as we showed.
Providing user-friendly ways to specify availability as a contract that the system will always respect during its operation will significantly improve the position of streaming systems in production environments.
Configuring availability in this way will probably impact resource utilization, performance overhead during normal operation, recovery time, and consistency.


\section{Load management, elasticity, \& reconfiguration}\label{sec:elasticity}

Due to the push-based nature of streaming inputs from external data sources, stream processors have no control over the rate of incoming events. Satisfying Quality of Service (QoS) under workload variations has been a long-standing research challenge in stream processing systems. 

To avoid performance degradation when input rates exceed system capacity, the stream processor needs to take actions that will ensure sustaining the load. One such action is \emph{load shedding}: temporarily dropping excess tuples from inputs or intermediate operators in the streaming execution graph.
Load shedding trades off result accuracy for sustainable performance and is suitable for applications with strict latency constraints that can tolerate approximate results.

When result correctness is more critical than low latency, dropping tuples is not an option. If the load increase is transient, the system can instead choose to reliably buffer excess data and process it later, once input rates stabilize. Several systems employ \emph{back-pressure}, a fundamental load management technique applicable to communication networks that involving producers and consumers. Nevertheless, to avoid running out of available memory during load spikes, \emph{load-aware scheduling} and rate control can be applied.  

A more recent approach that aims at satisfying QoS while guaranteeing result correctness under variable input load is \emph{elasticity}. Elastic stream processors are capable of adjusting their configuration and scaling their resource allocation in response to load. Dynamic scaling methods are applicable to both centralized and distributed settings. Elasticity not only addresses the case of increased load, but can additionally ensure no resources are left idle when the input load decreases.

Next, we review load shedding (Section~\ref{sec:load-shedding}), load-aware scheduling and flow control  (Section~\ref{sec:flow-control}), and elasticity techniques (Section~\ref{sec:elasticity-inner}). As in previous sections, we conclude with a discussion of 1st generation vs. modern and open problems.

\subsection{Load shedding}\label{sec:load-shedding}
Load shedding~\cite{tatbul2003load,tatbul2007staying,babcock2004load,Tu2006load} is the process of discarding data when input rates increase beyond system capacity. The system continuously monitors query performance and if an overload situation is detected, it selectively drops tuples according to a QoS specification. 
Load shedding is commonly implemented by a standalone component integrated with the stream processor. The load shedder continuously monitors input rates or other system metrics and can access information about the running query plan. Its main functionality consists of detecting overload (\emph{when} to shed load) and deciding what actions to take in order to maintain acceptable latency and minimize result quality degradation. These actions presume answering the questions of \emph{where} (in the query plan), \emph{how many}, and \emph{which} tuples to drop.


Detecting overload is a crucial task, as an incorrectly triggered shedding action can cause unnecessary result degradation. To facilitate the decision of \emph{when}, load shedding components rely on statistics gathered during execution.  The more knowledge a load shedder has about the query plan and its execution, the more accurate decisions it can make. For this reason, many stream processors restrict load shedding to a predefined set of operators, such as those that do not modify tuples, i.e.\ filter, union, and join~\cite{tatbul2003load,kang2003evaluating, das03approximate}. Other operator-restricted load shedding techniques target window operators~\cite{babcock2004load, tatbul2006window}, or even more specifically, query plans with \texttt{SUM} or \texttt{COUNT} sliding window aggregates~\cite{babcock2004load}.  An alternative, operator-independent approach, is to frame load shedding as a feedback control problem~\cite{Tu2006load}. The load shedder relies on a dynamic model that describes the relationship between average tuple delay (latency) and input rate. 

Once the load shedder has detected overload, it needs to perform the actual load shedding. This includes the decision of where in the query plan to drop tuples from, as well as which tuples and how many. 
The question of where is equivalent to placing special \emph{drop operators} in the best positions in the query plan. In general, drop operators can be placed at any location in the query plan, however, they are often placed at or near the sources. Dropping tuples early avoids wasting work but it might affect results of multiple queries if the stream processor operates on a shared query network. Alternatively, a load shedding road map (LSRM) can be used~\cite{tatbul2003load}. This is a pre-computed table that contains materialized load shedding plans, ordered by the amount of load shedding they will cause. 

The question of which tuples to drop is relevant when load shedding takes into account the \emph{semantic} importance of tuples with respect to results quality.
A \emph{random} dropping strategy has been applied to sliding window aggregate queries to provide approximate results  by inserting random sampling operators in the query plan~\cite{babcock2004load}.
\emph{Window-aware} load shedding~\cite{tatbul2006window} applies shedding to entire windows instead of individual tuples, while 
\emph{concept-driven} load shedding~\cite{katsipoulakis2018concept} is a semantic dropping strategy that selects tuples to discard based on the notion of concept drift.  

\subsection{Scheduling and flow control}\label{sec:flow-control}
When load bursts are transient and a temporary increase in latency is preferred to missing results, back-pressure and flow control can provide load management without sacrificing accuracy.
Flow control methods include buffering excess load, load-aware scheduling that prioritizes operators with the objective to minimize the backlog, regulating the transmission rate, and throttling the producer.
Flow control and back-pressure techniques do not consider application-level quality requirements, such as the semantic importance of input tuples. Their main requirement is availability of buffer space at the sources or intermediate operators and that any accumulated load is within the system capacity limits, so that it will be eventually possible to process the data backlog.

\stitle{Load-aware scheduling} tackles the overload problem by selecting the \emph{order} of operator execution and by adapting the \emph{resource allocation}. For instance, backlog can be reduced by dynamically selecting the order of executing filters and joins~\cite{avnur2000eddies,babu2004adaptive}. Alternatively, adaptive scheduling~\cite{Babcock2003Chain,Carney2003scheduling} modifies the allocation of resources given a static query plan.
The objective of load-aware scheduling strategies is to select an operator execution order that minimizes the total size of input queues in the system. The scheduler relies on knowledge about operator selectivities and processing costs. These statistics are either assumed to be known in advance, or need to be collected periodically during runtime. Operators are assigned priorities that reflect their potential to minimize intermediate results, and, consequently, the size of queues. 

\stitle{Back-pressure and flow control.} In a network of consumers and producers such as a streaming execution graph with multiple operators, back-pressure has the effect that all operators slow down to match the processing speed of the slowest consumer. If the bottleneck operator is far down the dataflow graph, back-pressure propagates to upstream operators, eventually reaching the data stream sources. To ensure no data loss, a persistent input message queue, such as Apache Kafka, and adequate storage space are required.

\begin{figure*}[ht]
    \centering
    \begin{subfigure}{.45\linewidth}
        \centering
        \includegraphics[width=.72\linewidth]{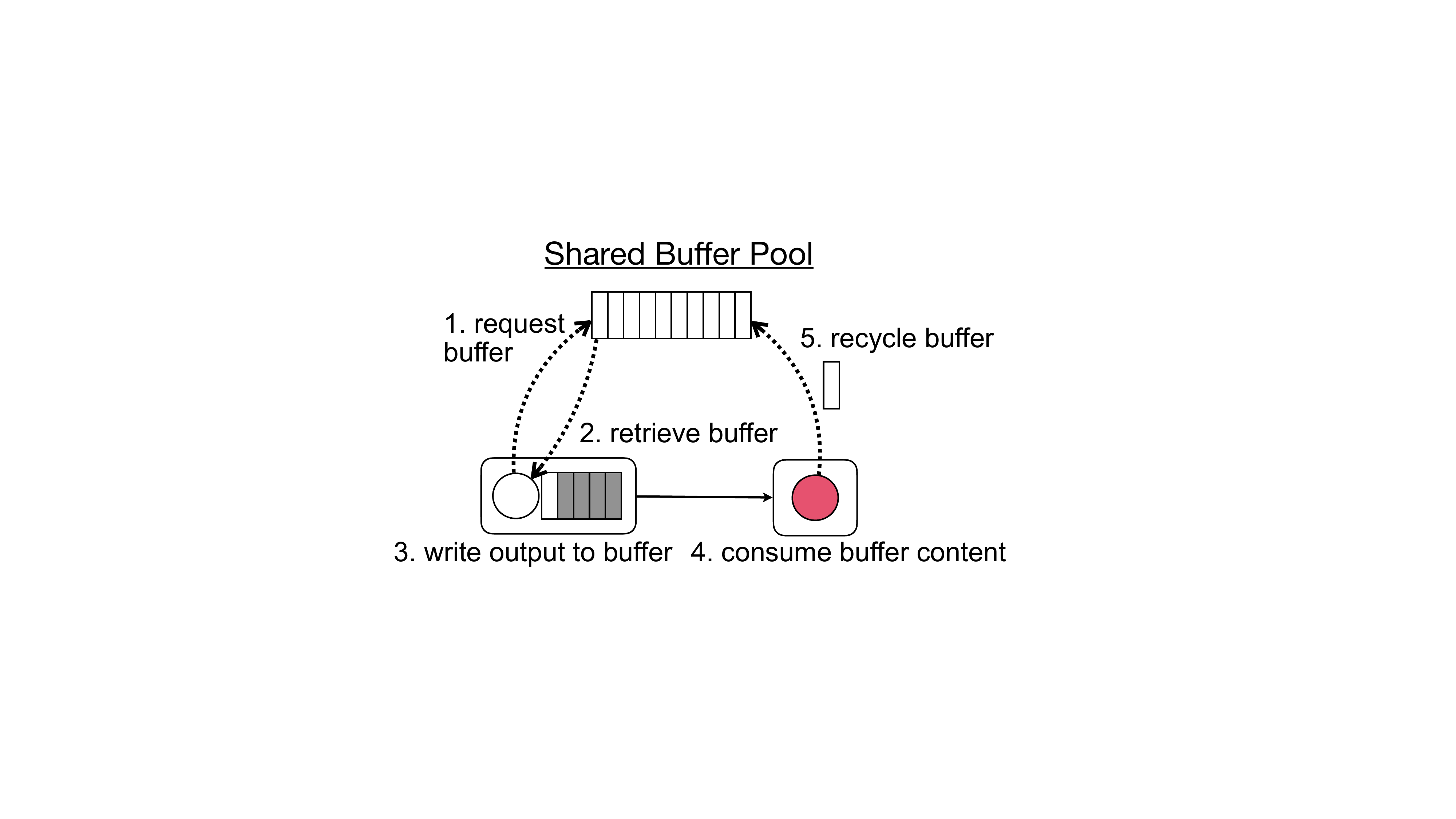}
        \caption{Local exchange}\label{fig:buffer-based-local}
    \end{subfigure}
    \begin{subfigure}{.48\linewidth}
        \centering
        \includegraphics[width=.95\linewidth]{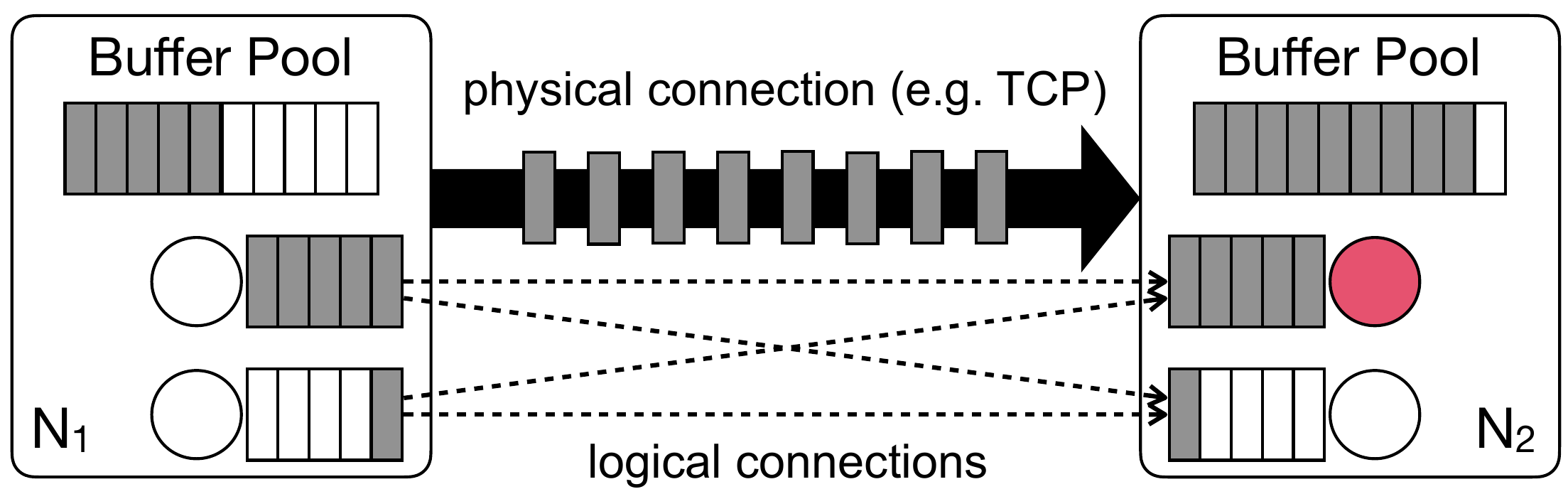}
        \caption{Remote exchange}\label{fig:buffer-based-remote}
    \end{subfigure}
    \caption{Buffer-based flow control.}
\end{figure*}

\emph{Buffer-based} back-pressure implicitly controls the flow of data via buffer availability. Considering a fixed amount of buffer space, a bottleneck operator will cause buffers to gradually fill up along its dataflow path.
Figure~\ref{fig:buffer-based-local} demonstrates buffer-based flow control when the producer and the consumer run on the same machine and share a buffer pool.
When a producer generates a result, it serializes it into an output buffer. If the producer and consumer run on the same machine and the consumer is slow, the producer might attempt to retrieve an output buffer when none will be available. The producer's processing rate will, thus, slow down according to the rate the consumer is recycling buffers back into the shared buffer pool.  The case when the producer and consumer are deployed on different machines and communicate via TCP is shown in Figure~\ref{fig:buffer-based-remote}. If no buffer is available on the consumer side, the TCP connection will be interrupted.  The producer can use a threshold to control how much data is in-flight and it is slowed down if it cannot put new data on the wire.

\begin{figure}[b]
  \centering
  \includegraphics[width=.9\linewidth]{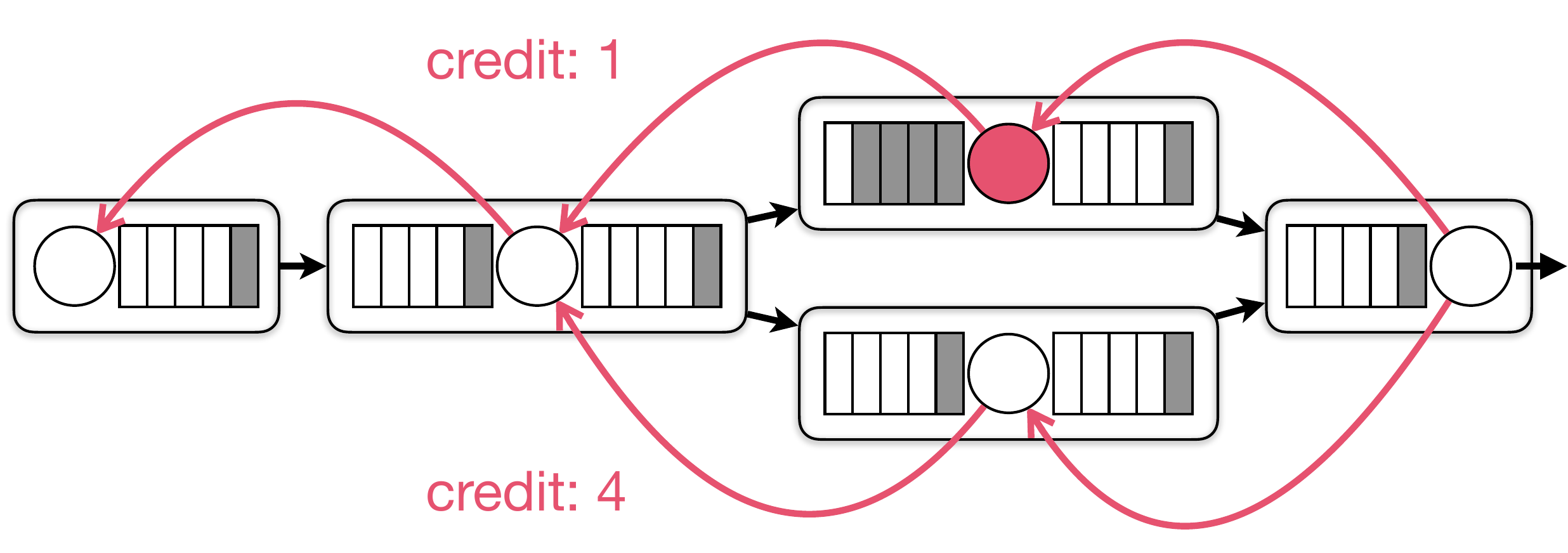}
   \caption{Credit-based flow control in a dataflow graph. Receivers regularly announce their credit upstream (gray and white squares indicate full and free buffers, respectively).}
   \label{fig:cfc}
\end{figure}



\emph{Credit-based flow control} (CFC)~\cite{Kung:994cfc} is a link-by-link, per virtual channel congestion control technique used in ATM network switches. 
In a nutshell, CFC uses a credit system to signal the availability of buffer space from receivers to senders. 
This classic networking technique turns out to be very useful for load management in modern, highly-parallel stream processors and is implemented in Apache Flink~\cite{CUSTOM:web/Flink}. Figure~\ref{fig:cfc} shows how the scheme works for a hypothetical dataflow. Parallel tasks are connected via virtual channels multiplexed over TCP connections. Each task informs its senders of its buffer availability via credit messages. This way, senders always know whether receivers have the required capacity to handle data messages. When the credit of a receiver drops to zero (or a specified threshold), back-pressure appears on its virtual channel. An important advantage of this per-channel flow control mechanism is that back-pressure is inflicted on pairs of communicating tasks only and does not interfere with other tasks sharing the same TCP connection. This is crucial in the presence of data skew where a single overloaded task could otherwise block the flow of data to all other downstream operator instances. On the downside, the additional credit announcement messages might increase end-to-end latency.


\subsection{Elasticity}\label{sec:elasticity-inner}
The approaches of load shedding and back-pressure are designed to handle workload variations in a \emph{statically provisioned} stream processor or application. 
Stream processors deployed on cloud environments or clusters can make use of a dynamic pool of resources. \emph{Dynamic scaling} or \emph{elasticity} is the ability of a stream processor to vary the resources available to a running computation in order to handle workload variations efficiently. Building an elastic streaming system requires a \emph{policy} and a \emph{mechanism}. The policy component implements a control algorithm that collects performance metrics and decides when and how much to scale. The mechanism effects the configuration change. It handles resource allocation, work re-assignment, and state migration, while guaranteeing result correctness. Table~\ref{tbl:scaling-comparison} summarizes the dynamic scaling capabilities and characteristics of elastic streaming systems.


\subsubsection{Elasticity policies}
A \emph{scaling policy} involves two individual decisions. First, it needs to detect the symptoms of an unhealthy computation and decide whether scaling is necessary. Symptom detection is a well-understood problem and can be addressed using conventional monitoring tools.
Second, the policy needs to identify the causes of exhibited symptoms (e.g. a bottleneck operator) and propose a scaling action. This is a challenging task which requires performance analysis and prediction.
It is common practice to place the burden of scaling decisions on application users who have to face conflicting incentives. They can either plan for the highest expected workload, possibly incurring high cost, or they can choose to be conservative and risk degraded performance.
Automatic scaling refers to scaling decisions transparently handled by the streaming system in response to load.
Commercial streaming systems that support automatic scaling include Google Cloud Dataflow~\cite{Kirpicho2016noshard}, Heron~\cite{KulkarniBF15}, and IBM System S~\cite{gedik2014elastic}, while DS2~\cite{Kalavri2018three}, Seep~\cite{castro2013integrating} and StreamCloud~\cite{gulisano2012streamcloud} are recent research prototypes.

In Table~\ref{tbl:scaling-comparison}, we categorize policies into \emph{heuristic} and \emph{predictive}. Heuristic policies rely on empirically predefined rules and are often triggered by thresholds or observed conditions while predictive policies make scaling decisions guided by analytical performance models.

Heuristic policy controllers gather coarse-grained metrics, such as CPU utilization, observed throughput, queue sizes, and memory utilization, to detect suboptimal scaling. CPU and memory utilization can be inadequate metrics for streaming applications deployed in cloud environments due to multi-tenancy and performance interference~\cite{rameshan2016hubbub}. StreamCloud~\cite{gulisano2012streamcloud} and Seep~\cite{castro2013integrating} try to mitigate the problem by separating user time and system time, but
preemption can make these metrics misleading. For example, high CPU usage caused by a task running on the same physical machine as a dataflow operator can trigger incorrect scale-ups (false positives) or prevent correct scale-downs (false negatives). Google Cloud Dataflow~\cite{Kirpicho2016noshard} relies on CPU utilization for scale-down decisions only but still suffers false negatives.
Dhalion~\cite{floratou2017dhalion} and IBM Streams~\cite{gedik2014elastic} also use congestion and back-pressure signals to identify bottlenecks. These metrics are helpful for identifying bottlenecks but they cannot detect resource over-provisioning.

\setlength{\tabcolsep}{4pt}
\begin{table*}
\smaller\centering
\caption{Elasticity policies and mechanisms in streaming systems}
\label{tbl:scaling-comparison}
\begin{tabular}{p{0.15\textwidth}
                        p{0.06\textwidth}
                        p{0.06\textwidth}
                        p{0.06\textwidth}
                        p{0.06\textwidth}
                        p{0.06\textwidth}
                        p{0.06\textwidth}
                        p{0.06\textwidth}
                        p{0.06\textwidth}
                        p{0.08\textwidth}
                        }
\hline
\multicolumn{1}{c}{\textbf{System}} &
\multicolumn{2}{c}{\textbf{Policy}} &
\multicolumn{2}{c}{\textbf{Objective}} &
\multicolumn{3}{c}{\textbf{Reconfiguration}} &
\multicolumn{2}{c}{\textbf{State Migration}}
\\
\rowcolor{white}
& \multicolumn{1}{c}{Heuristic}
& \multicolumn{1}{c}{Predictive}

& \multicolumn{1}{c}{Latency}
& \multicolumn{1}{c}{Throughput}

& \multicolumn{1}{c}{Stop-and-Restart}
& \multicolumn{1}{c}{Partial Pause}
& \multicolumn{1}{c}{Live} 

& \multicolumn{1}{c}{At-Once}
& \multicolumn{1}{c}{Progressive} \\
\hline

Borealis~\cite{abadi2005design}
& \multicolumn{1}{a}{\checkmark}
& \multicolumn{1}{a}{}
& \multicolumn{1}{b}{\checkmark}
& \multicolumn{1}{b}{\checkmark}
& \multicolumn{3}{a}{n/a}
& \multicolumn{2}{b}{n/a}
\\
StreamCloud~\cite{gulisano2012streamcloud} 
& \multicolumn{1}{a}{\checkmark}
& \multicolumn{1}{a}{}
& \multicolumn{1}{b}{}
& \multicolumn{1}{b}{\checkmark}
& \multicolumn{1}{a}{}
& \multicolumn{1}{a}{\checkmark}
& \multicolumn{1}{a}{}
& \multicolumn{1}{b}{\checkmark}
& \multicolumn{1}{b}{}
\\
Seep~\cite{castro2013integrating} 
& \multicolumn{1}{a}{\checkmark}
& \multicolumn{1}{a}{}
& \multicolumn{1}{b}{\checkmark}
& \multicolumn{1}{b}{\checkmark}
& \multicolumn{1}{a}{}
& \multicolumn{1}{a}{\checkmark}
& \multicolumn{1}{a}{}
& \multicolumn{1}{b}{\checkmark}
& \multicolumn{1}{b}{}
\\
IBM Streams~\cite{gedik2014elastic} 
& \multicolumn{1}{a}{\checkmark}
& \multicolumn{1}{a}{}
& \multicolumn{1}{b}{}
& \multicolumn{1}{b}{\checkmark}
& \multicolumn{1}{a}{}
& \multicolumn{1}{a}{\checkmark}
& \multicolumn{1}{a}{}
& \multicolumn{1}{b}{\checkmark}
&  \multicolumn{1}{b}{}
\\
FUGU~\cite{heinze14latency,heinze2014auto-scaling} 
& \multicolumn{1}{a}{\checkmark}
& \multicolumn{1}{a}{}
& \multicolumn{1}{b}{}
& \multicolumn{1}{b}{\checkmark}
& \multicolumn{1}{a}{}
& \multicolumn{1}{a}{\checkmark}
& \multicolumn{1}{a}{}
& \multicolumn{1}{b}{\checkmark}
& \multicolumn{1}{b}{}
\\
Nephele~\cite{lohrmann2015elastic} 
& \multicolumn{1}{a}{}
& \multicolumn{1}{a}{\checkmark}
& \multicolumn{1}{b}{\checkmark}
& \multicolumn{1}{b}{}
& \multicolumn{1}{a}{}
& \multicolumn{1}{a}{}
& \multicolumn{1}{a}{}
& \multicolumn{1}{b}{}
& \multicolumn{1}{b}{}
\\
DRS~\cite{fu2017DRS} 
& \multicolumn{1}{a}{}
& \multicolumn{1}{a}{\checkmark}
& \multicolumn{1}{b}{\checkmark}
& \multicolumn{1}{b}{}
& \multicolumn{1}{a}{}
& \multicolumn{1}{a}{}
& \multicolumn{1}{a}{}
& \multicolumn{1}{b}{}
& \multicolumn{1}{b}{}
\\
MPC~\cite{dematteis2017elastic} 
& \multicolumn{1}{a}{}
& \multicolumn{1}{a}{\checkmark}
& \multicolumn{1}{b}{\checkmark}
& \multicolumn{1}{b}{}
& \multicolumn{1}{a}{}
& \multicolumn{1}{a}{\checkmark}
& \multicolumn{1}{a}{}
& \multicolumn{1}{b}{\checkmark}
& \multicolumn{1}{b}{}
\\
CometCloud~\cite{tolosana2017feedback} 
& \multicolumn{1}{a}{}
& \multicolumn{1}{a}{\checkmark}
& \multicolumn{1}{b}{\checkmark}
& \multicolumn{1}{b}{}
& \multicolumn{1}{a}{}
& \multicolumn{1}{a}{}
& \multicolumn{1}{a}{\checkmark}
& \multicolumn{2}{b}{n/a}
\\
Chronostream~\cite{wu2015Chronostream} 
& \multicolumn{2}{a}{n/a}
& \multicolumn{2}{b}{n/a}
& \multicolumn{1}{a}{}
& \multicolumn{1}{a}{}
& \multicolumn{1}{a}{\checkmark}
& \multicolumn{1}{b}{\checkmark}
& \multicolumn{1}{b}{}
\\
ACES~\cite{amini2006adaptive} 
& \multicolumn{1}{a}{}
& \multicolumn{1}{a}{\checkmark}
& \multicolumn{1}{b}{\checkmark}
& \multicolumn{1}{b}{\checkmark}
& \multicolumn{3}{a}{n/a}
& \multicolumn{2}{b}{n/a}
\\
Stella~\cite{xu2016stela} 
& \multicolumn{1}{a}{\checkmark}
& \multicolumn{1}{a}{}
& \multicolumn{1}{b}{}
& \multicolumn{1}{b}{\checkmark}
& \multicolumn{1}{a}{}
& \multicolumn{1}{a}{}
& \multicolumn{1}{a}{}
& \multicolumn{1}{b}{}
& \multicolumn{1}{b}{}
\\ 
Google~Dataflow~\cite{Kirpicho2016noshard}
& \multicolumn{1}{a}{\checkmark}
& \multicolumn{1}{a}{}
& \multicolumn{1}{b}{\checkmark}
& \multicolumn{1}{b}{\checkmark}
& \multicolumn{1}{a}{}
& \multicolumn{1}{a}{}
& \multicolumn{1}{a}{}
& \multicolumn{1}{b}{}
& \multicolumn{1}{b}{}
\\ 
Dhalion~\cite{floratou2017dhalion}
& \multicolumn{1}{a}{\checkmark}
& \multicolumn{1}{a}{}
& \multicolumn{1}{b}{}
& \multicolumn{1}{b}{\checkmark}
& \multicolumn{1}{a}{\checkmark}
& \multicolumn{1}{a}{}
& \multicolumn{1}{a}{}
& \multicolumn{1}{b}{\checkmark}
& \multicolumn{1}{b}{}
\\ 
DS2~\cite{Kalavri2018three}
& \multicolumn{1}{a}{}
& \multicolumn{1}{a}{\checkmark}
& \multicolumn{1}{b}{}
& \multicolumn{1}{b}{\checkmark}
& \multicolumn{1}{a}{\checkmark}
& \multicolumn{1}{a}{}
& \multicolumn{1}{a}{}
& \multicolumn{1}{b}{\checkmark}
& \multicolumn{1}{b}{}
\\ 
Spark Streaming~\cite{Zaharia:2013:DSF:2517349.2522737,armbrust2018structured}
& \multicolumn{1}{a}{\checkmark}
& \multicolumn{1}{a}{}
& \multicolumn{1}{b}{}
& \multicolumn{1}{b}{\checkmark}
& \multicolumn{1}{a}{\checkmark}
& \multicolumn{1}{a}{}
& \multicolumn{1}{a}{}
& \multicolumn{1}{b}{\checkmark}
& \multicolumn{1}{b}{}
\\
Megaphone~\cite{hoffmann19megaphone}
& \multicolumn{1}{a}{}
& \multicolumn{1}{a}{}
& \multicolumn{1}{b}{}
& \multicolumn{1}{b}{}
& \multicolumn{1}{a}{}
& \multicolumn{1}{a}{\checkmark}
& \multicolumn{1}{a}{}
& \multicolumn{1}{b}{}
& \multicolumn{1}{b}{\checkmark}
\\
Turbine~\cite{mei20turbine}
& \multicolumn{1}{a}{\checkmark}
& \multicolumn{1}{a}{}
& \multicolumn{1}{b}{}
& \multicolumn{1}{b}{\checkmark}
& \multicolumn{1}{a}{\checkmark}
& \multicolumn{1}{a}{}
& \multicolumn{1}{a}{}
& \multicolumn{1}{b}{\checkmark}
& \multicolumn{1}{b}{}
\\
Rhino~\cite{del2020rhino}
& \multicolumn2{a}{n/a}
& \multicolumn{2}{b}{n/a}
& \multicolumn{1}{a}{}
& \multicolumn{1}{a}{\checkmark}
& \multicolumn{1}{a}{}
& \multicolumn{1}{b}{\checkmark}
& \multicolumn{1}{b}{}
\\
\hline

\end{tabular}
\end{table*}

  
Predictive policy controllers build an analytical performance model of the streaming system and formulate the scaling problem as a set of mathematical functions. Predictive approaches include queuing theory~\cite{lohrmann2015elastic,fu2017DRS,tolosana2017feedback,fu2017DRS}, control theory~\cite{dematteis2017elastic,amini2006adaptive,khoshkbarforoushha2017Elasticity}, and instrumentation-driven linear performance models~\cite{Kalavri2018three}. Thanks to their closed-form analytical formulation, predictive policies are capable of making multi-operator decisions in one step.

\subsubsection{Elasticity mechanisms}

Elasticity mechanisms are concerned with realizing the actions indicated by the policy. 
They need to ensure correctness and low-latency redistribution of accumulated state when effecting a reconfiguration. To ensure correctness, many streaming systems rely on the fault-tolerance mechanism to provide reconfiguration capabilities. When adding new workers to a running computation, the mechanism needs not only re-assign work to them but also migrate any necessary state these new workers will now be in charge of. 
Elasticity mechanisms need to complete a reconfiguration as quickly as possible and at the same time minimize performance disruption. We review the main methods for state redistribution, reconfiguration, and state transfer next. We focus on systems with embedded state, as reconfiguration mechanisms are significantly simplified when state is external.

\stitle{State redistribution.}
State redistribution must preserve key semantics, so that existing state for a particular key and all future events with this key are routed to the same worker. For that purpose, most systems use hashing methods. \emph{Uniform hashing} evenly distributes keys across parallel tasks. It is fast to compute and requires no routing state but might incur high migration cost. When a new node is added, state is shuffled across existing and new workers. It also causes random I/O and high network communication. Thus, it is not particularly suitable for adaptive applications. \emph{Consistent hashing} and variations are more often preferred. Workers and keys are mapped to multiple points on a ring using multiple random hash functions. Consistent hashing ensures that state is not moved across workers that are present before and after the migration. When a new worker joins, it becomes responsible for data items from multiple of the existing nodes. When a worker leaves, its key space is distributed over existing workers. 
Apache Flink~\cite{carbone2015apache} uses a variation of consistent hashing in which state is organized into \emph{key groups} and those are mapped to parallel tasks as ranges. On reconfiguration, reads are sequential within each key group, and often across multiple key groups. The metadata of key group to task assignments are small and it is sufficient to store key-group range boundaries. The number of key groups limits the maximum number of parallel tasks to which keyed state can be scaled.

Hashing techniques are simple to implement and do not require storing any routing state, however, they do not perform well under skewed key distributions. \emph{Hybrid partitioning}~\cite{gedik2014partitioning} combines consistent hashing and an explicit mapping to generate a compact hash function that provides load balance in the presence of skew. The main idea is to track the frequencies of the partitioning key values and treat normal keys and popular keys differently. The mechanism uses the lossy counting algorithm~\cite{manku2002approximate}  in a sliding window setting to estimate heavy hitters, as keeping exact counts would be impractical for large key domains.

\stitle{Reconfiguration strategy.}
Regardless of the re-partitioning strategy used, if the elasticity policy makes a decision to change an application's resources, the mechanism will have to transfer some amount of state across workers on the same or different physical machines.

The \emph{stop-and-restart} strategy halts the computation, takes a state snapshot of all operators, and then restarts the application with the new configuration. Even though this mechanism is simple to implement and it trivially guarantees correctness, it unnecessary stalls the entire pipeline even if only one or few operators need to be rescaled. As shown in Table~\ref{tbl:scaling-comparison}, this strategy is very common in modern systems.

\begin{figure}[t]
  \centering
  \includegraphics[width=.9\linewidth]{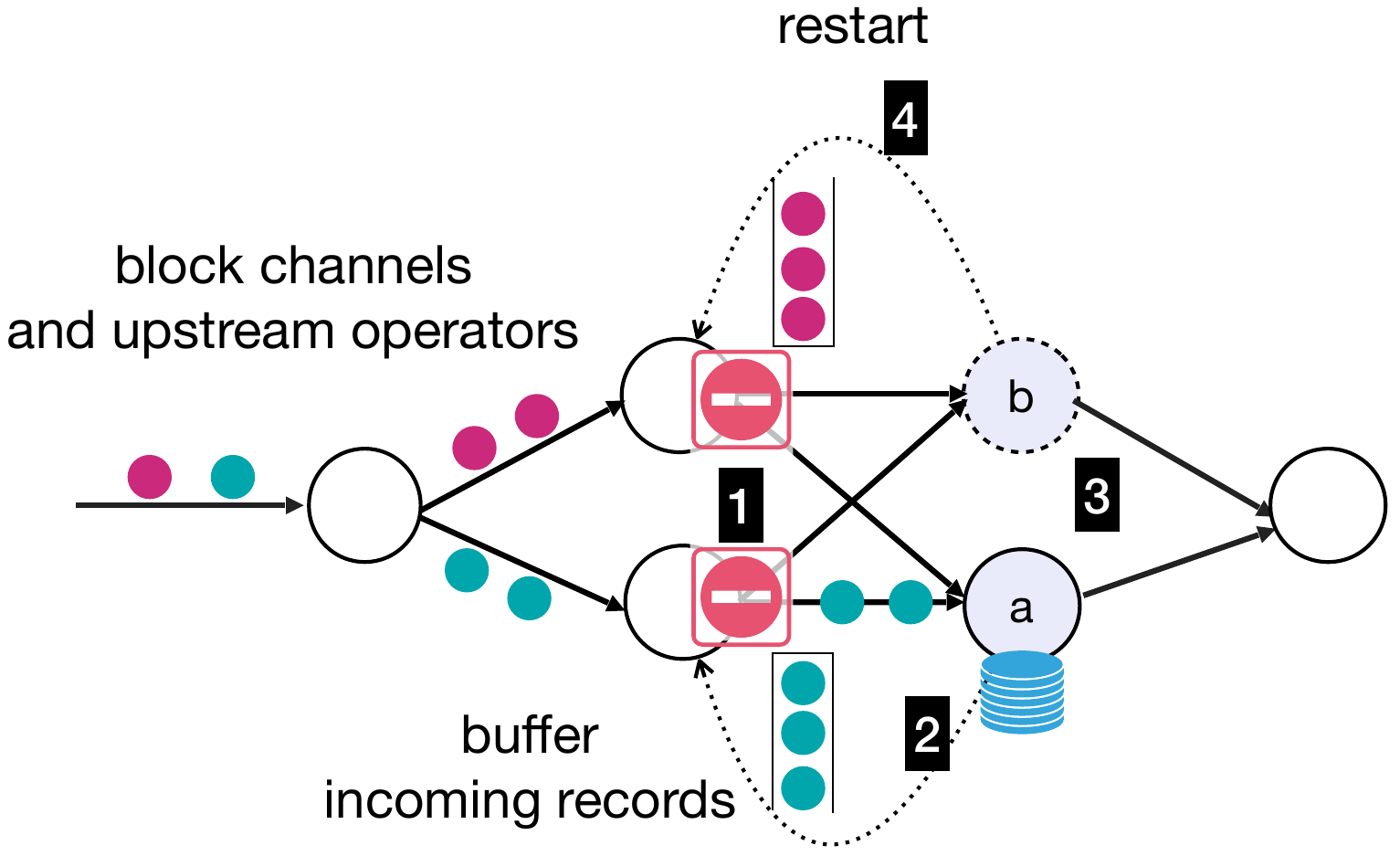}
   \caption{An example of the partial-pause-and-restart protocol. To move state from operator $a$ to $b$, the mechanism executes the following steps: (1) Pause $a$'s upstream operators, (2) extract state from $a$, (3) load state into $b$, and (4) send a restart signal from $b$ to upstream operators.}
   \label{fig:partial-pause-and-restart}
\end{figure}

\emph{Partial pause and restart}, introduced by FLUX~\cite{shah2003flux}, is a less disruptive strategy that only blocks the affected dataflow subgraph temporarily. The affected subgraph contains the operator to be scaled, as well as upstream channels and upstream operators. Figure~\ref{fig:partial-pause-and-restart} shows an example of the protocol. To migrate state from operator $a$ to operator $b$, the mechanism will execute the following steps:
(1) First, it \emph{pauses} $a$'s upstream operators and stops pushing tuples to $a$. Paused operators start buffering input tuples in their local buffers. operator $a$ continues processing tuples in its buffers until they are empty. (2) Once $a$'s buffers are empty, it extracts its state and sends it to operator $b$. (3) Operator $b$ loads the state and (4) sends a \emph{restart} signal to upstream operators. Once upstream operators receive the signal they can start processing tuples again.

\begin{figure}[b]
  \centering
  \includegraphics[width=.9\linewidth]{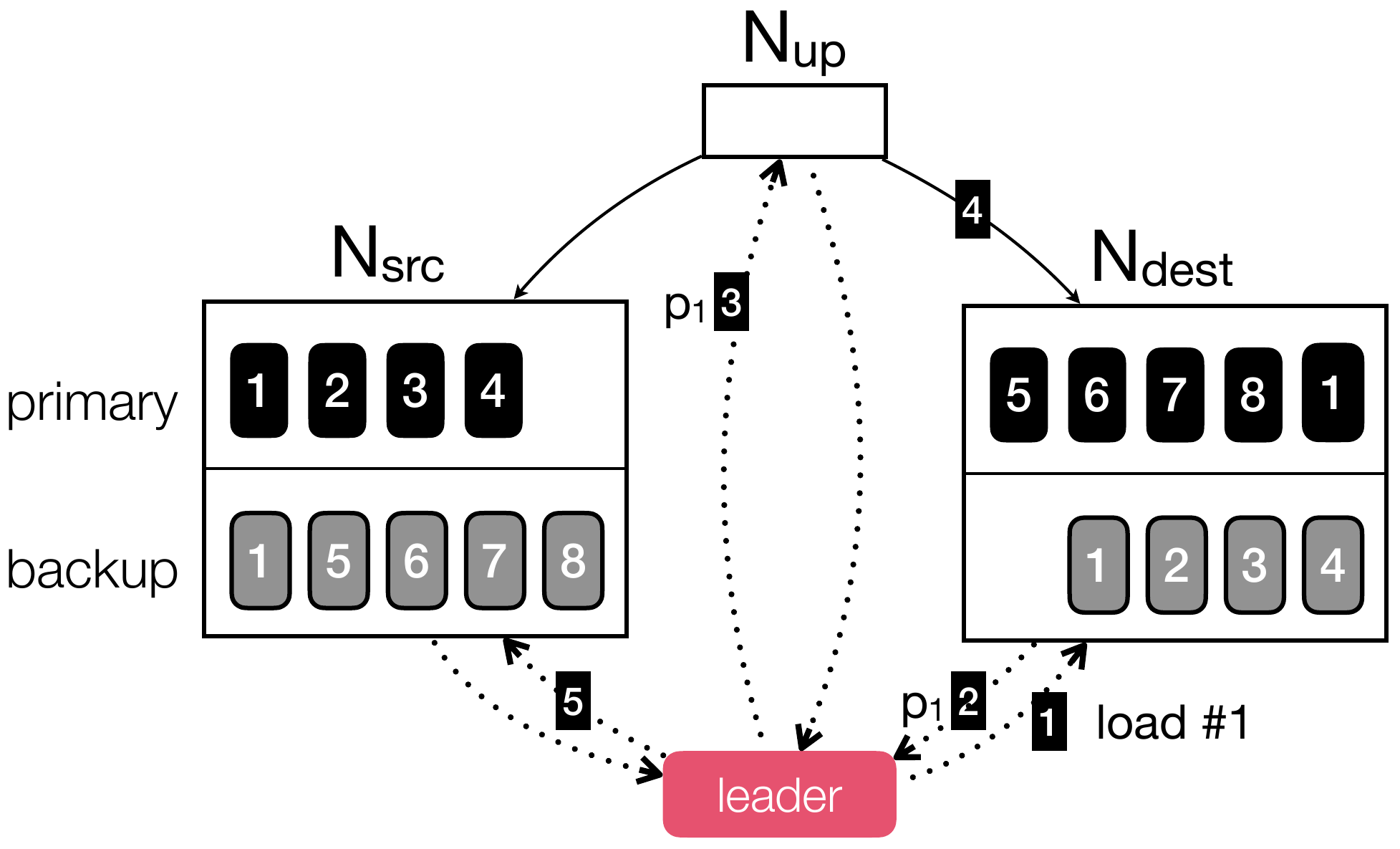}
   \caption{An example of the proactive replication protocol. To move slice \#1 from $N_{src}$ to $N_{dest}$, the mechanism executes the following steps: (1) the leader instructs $N_{dest}$ to load slice \#1, (2) $N_{dest}$ loads slice \#1 and sends ack to the leader, (3) the leader notifies upstream operators to replay events, (4) upstream start rerouting events to $N_{dest}$, (5) the leader notifies $N_{src}$ that the transfer is complete and $N_{src}$ moves slice \#1 to the backup group.}
   \label{fig:proactive}
\end{figure}

The \emph{pro-active replication} strategy maintains state backup copies in multiple nodes so that reconfiguration can be performed in a nearly live manner when needed. The state is organized into smaller partitions, each of which can be transferred independently. Each node has a set of primary state slices and a set of secondary state slices. Figure~\ref{fig:proactive} shows an example of the protocol as implemented by ChronoStream~\cite{wu2015Chronostream}. 

\stitle{State transfer.} Another important decision to make when migrating state from one worker to another is whether the state is moved \emph{all-at-once} or in a \emph{progressive} manner. If a large amount of state needs to be transferred, moving it in one operation might cause high latency during re-configuration. 
Alternatively, \emph{progressive} migration~\cite{hoffmann19megaphone} moves state in smaller pieces and flattens latency spikes by interleaving state transfer with processing. On the downside, progressive state migration might lead to longer migration duration.

\subsection{1st generation vs. 2nd generation}

Comparing early to modern approaches, we make the following observations. While load shedding was popular among early stream processors, modern systems do not favor the approach of degrading results quality anymore. Another important difference is that load management approaches in 1st generation systems used to affect the execution of multiple queries as they formed a shared dataflow plan (cf. Section~\ref{sec:preliminaries}). Queries in modern systems are typically executed as independent jobs, thus, back-pressure on a certain query will not affect the execution of other queries running on the same cluster. Scaling down is a quite recent requirement that was not a matter of concern before cloud deployments. 
The dependence on persistent queues for providing correctness guarantees is another recent characteristic, mainly required by systems employing back-pressure. Finally, while early load shedding and load-aware scheduling techniques assume a limited set of operators whose properties and characteristics are stable throughout execution, modern systems implement general load management methods that are applicable even if cost and selectivity vary or are unknown.

\subsection{Open Problems}
Adaptive scheduling methods have so far been studied in the context of simple query plans with operators whose selectivities and costs are fixed and known. It is unclear whether these methods generalize to arbitrary plans, operators with UDFs, general windows, and custom joins. Load-aware scheduling can further cause starvation and increased per-tuple latency, as low-priority operators with records in their input buffers would need to wait a long time during bursts. Finally, existing methods are restricted to streams that arrive in timestamp order and do not support out-of-order or delayed events.

Re-configurable stream processing is a quite recent research area, where stream processors are designed to not only be capable of adjusting their resource allocation but other elements of their runtime as well.
Elasticity, the ability of a stream processor to dynamically adjust resource allocation can be considered as a special case of re-configuration. Others include code updates for bug fixes, version upgrades, or business logic changes, execution plan switching, dynamic scheduling and operator placement, as well as skew and straggler mitigation. So far, each of the aforementioned re-configuration scenarios have been largely studied in isolation. To provide general re-configuration and self-management, future systems will need to take into account how optimizations interact with each other.

\section{Conclusion}\label{sec:conclusion}
While early streaming systems strove to extend relational execution engines with time-based window processing, modern systems have evolved significantly in terms of architecture and capabilities. Table~\ref{tab:evolution-streaming} summarizes the evolution of major streaming system aspects over the last three decades.

While approximate results were mainstream in early systems, modern systems have primarily focused on results correctness and have largely rejected the notion of approximation.
In terms of languages, modern systems favor general-purpose programming languages, however, we recently witness a trend to return to extensions for streaming SQL~\cite{streamsql}. Over the years, execution has also gradually transitioned from mainly centralized to mainly distributed, exploiting data, pipeline, and task parallelism. At the same time, most modern systems construct independent execution plans per query and apply little optimization and sharing.



Regarding time, order, and progress, many of the inventions of the past proved to have survived the test of time, since they continue to hold a place in modern streaming systems.
Especially Millwheel and the Google Dataflow Model popularized punctuations, watermarks, the out-of-order architecture, and triggers for revision processing. Streaming state management witnessed a major shift, from specialized in-memory synopses to large partitioned and persistent state supported today. As a result, fault tolerance and high availability also shifted towards passive replication and exactly-once processing. Finally, load management approaches have transitioned from load shedding and scheduling methods to elasticity and backpressure coupled with persistent inputs.

In state management we identify the most radical changes seen in data streaming so far. The most obvious advances relate to the scalability of state and long-term persistence in unbounded executions. Yet, today's systems have invested thoroughly in providing transactional guarantees that are in par with those modern database management systems can offer today. Transactional stream processing has pivoted data streaming beyond the use for data analytics and has also opened new research directions in terms of efficient methods for backing and accessing state that grows in unbounded terms. Stream state and compute are gradually being decoupled and this allows for better optimizations, wider interoperability with storage technologies as well as novel semantics for shared and external state having stream processors as the backbone of modern continuous applications and live scalable data services.

We believe the road ahead is still long for streaming systems. Emerging streaming applications in the areas of Cloud services~\cite{akhter2019stateful,Goldstein2020providing}, machine learning~\cite{garefalakis2019neptune,meldrum2019arcon}, and streaming graph analytics~\cite{Besta2020practice,abbas2018streaming} present new requirements and are already shaping the key characteristics of the future generation of data stream technology. We expect systems to evolve further and exploit next-generation hardware~\cite{zeuch2019analyzing,zhang2020hardware}, focus on transactions and iteration support, improve their reconfiguration capabilities, and take state management a step further by leveraging workload-aware backends~\cite{kalavri2020support}, shared state and versioning.

\bibliographystyle{abbrv}
\interlinepenalty=10000
\bibliography{references}

\balance

\end{document}